%% file: ms.tex
\documentclass[preprint]{aastex}
\slugcomment{\today{}}

\def\ngavg{\bar{n}_g}

\def\Navg{\langle N\rangle_M}
\def\Mmin{M_{\rm min}}
\def\NNm1{\langle N(N-1) \rangle}

\def\hMpc{h^{-1}{\rm Mpc}}
\def\Ncen{N_{\rm cen}}
\def\Nsat{N_{\rm sat}}
\def\Ncenavg{\langle N_{\rm cen}\rangle_M}
\def\Nsatavg{\langle N_{\rm sat}\rangle_M}

\def\Msun{M_\odot}
\def\hMsun{M_\odot}
\def\sigM{\sigma_{\log\,M}}
\def\sigMg{\sigma_{\log\,M_g}}
\def\sigMc{\sigma_{\log\,M_c}}
\def\sigMs{\sigma_{\log\,M_s}}
\def\Mgal{M_{\rm gal}}
\def\Mr{M_r}

\begin{document}

\shortauthors{Zheng et al.}
\shorttitle{Theoretical Models of the HOD}

\title{Theoretical Models of the Halo Occupation Distribution: Separating
Central and Satellite Galaxies}
\author{
Zheng Zheng,\altaffilmark{1,2,3}
Andreas A. Berlind,\altaffilmark{4}
David H. Weinberg,\altaffilmark{1}
Andrew J. Benson,\altaffilmark{5}
Carlton M. Baugh,\altaffilmark{6}
Shaun Cole,\altaffilmark{6}
Romeel Dav\'e,\altaffilmark{7}
Carlos S. Frenk,\altaffilmark{6}
Neal Katz,\altaffilmark{8}
and Cedric G. Lacey\altaffilmark{6}
}
\altaffiltext{1}{Department of Astronomy, The Ohio State University,
Columbus, OH 43210; zhengz@astronomy.ohio-state.edu, 
dhw@astronomy.ohio-state.edu.}
\altaffiltext{2}{Institute for Advanced Study, Einstein Drive, Princeton, 
NJ 08540; zhengz@ias.edu}
\altaffiltext{3}{Hubble Fellow}
\altaffiltext{4}{Center for Cosmology and Particle Physics, New York 
University, New York, NY 10003; aberlind@cosmo.nyu.edu.}
\altaffiltext{5}{Department of Astronomy, California Institute of Technology,
Pasadena, CA 91125; abenson@astro.caltech.edu.}
\altaffiltext{6}{Institute for Computational Cosmology, University of Durham, 
Durham DH1 3LE, UK; c.m.baugh@durham.ac.uk, shaun.cole@durham.ac.uk, 
c.s.frenk@durham.ac.uk, cedric.lacey@durham.ac.uk.}
\altaffiltext{7}{Steward Observatory, University of Arizona, Tucson, AZ 85721; 
rad@as.arizona.edu.}
\altaffiltext{8}{Department of Physics and Astronomy, University of 
Massachusetts, Amherst, MA 01003; nsk@kaka.phast.umass.edu.}

\begin{abstract}
The halo occupation distribution (HOD) describes the relation between
galaxies and dark matter at the level of individual dark matter halos.
The properties of galaxies residing at the centers of halos differ from those
of satellite galaxies because of differences in their formation histories.
Using a smoothed particle hydrodynamics (SPH) simulation and a semi-analytic
(SA) galaxy formation model, we examine the separate contributions of central
and satellite galaxies to the HOD, more specifically to the probability
$P(N|M)$ that a halo of virial mass $M$ contains $N$ galaxies of a particular
class.  In agreement with earlier results for dark matter subhalos, we find
that the mean occupation function $\Navg$ for galaxies above a baryonic mass
threshold can be approximated by a step function for central galaxies plus a
power law for satellites, and that the distribution of satellite numbers is
close to Poisson at fixed halo mass.  Since the number of central galaxies is
always zero or one, the width of $P(N|M)$ is narrower than a Poisson
distribution at low $N$ and approaches Poisson at high $N$.  For galaxy samples
defined by different baryonic mass thresholds, there is a nearly linear
relation between the minimum halo mass $\Mmin$ required to host a central
galaxy and the mass $M_1$ at which an average halo hosts one satellite,
with $M_1 \approx 14\Mmin$ (SPH) or $M_1 \approx 18\Mmin$ (SA).  The stellar
population age of central galaxies correlates with halo mass, and
this correlation explains much of the age-dependence of the galaxy HOD.
The mean occupation number of young galaxies exhibits a local minimum at
$M \sim 10\Mmin$ where halos are too massive to host a young central galaxy
but not massive enough to host satellites.  Using the SA model, we show that
the conditional galaxy mass function at fixed halo mass cannot be described
by a Schechter function because central galaxies produce a ``bump'' at high
masses.  We suggest parameterizations for the HOD and the conditional
luminosity function that can be used to model observed galaxy clustering.
Many of our predictions are in good agreement with recent results inferred
from clustering in the Sloan Digital Sky Survey.
\end{abstract}


\section{Introduction}

The Halo Occupation Distribution (HOD) has emerged as a powerful framework for 
describing galaxy bias and modeling galaxy clustering (e.g., 
\citealt{Ma00,Peacock00,Seljak00,Scoccimarro01,Berlind02}). It characterizes 
the bias between galaxies and mass in terms of the probability distribution
$P(N|M)$ that a halo of virial mass $M$ contains $N$ galaxies of a given type,
together with relative spatial and velocity distributions of galaxies and dark
matter within halos. 
The HOD as a function of galaxy type (defined by luminosity, color, morphology,
etc.) is a fundamental prediction of galaxy formation theory
(e.g., \citealt{Kauffmann97,Kauffmann99,Benson00,White01,Yoshikawa01,Berlind03,
Kravtsov04}), which can be tested by deriving the HOD empirically from
observed clustering.  Berlind et al. (2003, hereafter B03) showed good
agreement between the predictions of smoothed particle hydrodynamics
(SPH) simulations and semi-analytic (SA) calculations for the same
cosmological model, and Kravtsov et al.\ (2004, hereafter K04) showed
that the HOD of substructures in high resolution N-body simulations has
many of the same features.  In this paper we extend the analysis of B03
by separately examining the contributions of central and satellite
galaxies to the HOD, following the approach of K04.

The central galaxies of halos are treated distinctly in SA models of
galaxy formation (e.g., \citealt{White91,Kauffmann93,Cole94,Avila98,
Kampen99,Somerville99}).  Central galaxies accrete most or all of a halo's
cooling gas.  In a merger between two halos, the most massive progenitor
galaxy becomes the central galaxy of the merged halo, and other,
satellite galaxies may merge with it after being dragged in by
dynamical friction.  In hydrodynamic simulations 
(e.g., \citealt{Cen92,Katz92,Evrard94,Pearce99,White01,Yoshikawa01})
there is no explicitly separate treatment of central and satellite
galaxies, but the central objects of halos emerge as a distinct
class, more massive and usually older than other galaxies in the same
halo (B03), in qualitative agreement with SA predictions.
The central galaxies in SPH simulations lie close to the most 
strongly bound dark matter particles and move slowly relative to the 
halo center-of-mass, while the radial profiles and velocity dispersions
of satellites are similar to those of dark matter.
B03 show that SPH and SA predictions of $P(N|M)$ are remarkably
similar provided one chooses mass thresholds that yield the same
galaxy number density.  They further show that the two methods
predict similar dependences of $P(N|M)$ on galaxy baryonic mass and
stellar population age.

K04 show that the description of $P(N|M)$ for subhalos in N-body 
simulations simplifies considerably if one distinguishes the contributions
of central and satellite substructures.  (We will use the terms
substructure and subhalo interchangeably.)  For a subhalo sample limited by
maximum circular velocity, the mean occupation function
$\Navg \equiv \sum_{N=0}^\infty N P(N|M)$ is well approximated by
the sum of a step function for the central substructure, 
$\Ncenavg = \Theta(M-\Mmin)$, where $\Theta(x)=0$ if 
$x<0$ and $\Theta(x)=1$ if $x\ge0$, and a power-law for the satellites,
$\Nsatavg=(M/M_1)^\alpha$, with $\alpha \approx 1$.  The resulting shape,
with a cutoff, a plateau at low $M$, and a power law at high $M$,
is similar to that found for SPH and SA model galaxies (B03).
More importantly, if one assumes that the distribution of $\Nsat$
with respect to the mean $\Nsatavg$ is a Poisson distribution,
then the model naturally explains the transition from a 
sub-Poisson width at low $\Navg$ (where the contribution of the 
central galaxy dominates) to a Poisson width at high $\Navg$ (where
the satellites dominate).  This transition is a common feature in 
all of the SA and SPH calculations mentioned above, and the sub-Poisson
fluctuations at low $\Navg$ play a crucial role in shaping the 
galaxy 2-point correlation function \citep{Benson00}.
\cite{Guzik02} also distinguish central and satellite galaxies in
their HOD modeling of galaxy-galaxy lensing, but they focus on
samples defined by 
bins of luminosity (rather than thresholds) and therefore
adopt a different parameterization.
They show that the SA models of \cite{Kauffmann99} predict a $\Navg$
for a narrow luminosity bin that is well approximated by a
Gaussian in $\log M$ for central galaxies and a power law for satellites.

Here we apply the K04 approach to the SPH and SA galaxy populations 
studied by B03.  We consider samples limited by thresholds in baryonic
mass, which should be similar to observational samples limited by
luminosity thresholds, particularly for observations at longer wavelengths.  
We also examine samples divided on the basis
of stellar population age, which should be comparable to observational
samples divided by color or spectral type.  Distinguishing central
and satellite galaxies proves especially valuable in understanding the
behavior of these ``red'' and ``blue'' galaxy HODs.  We present simple
parameterizations of the predicted HODs, which can guide efforts
to infer the HOD from observed galaxy clustering.  We also present
results for the conditional luminosity function of galaxies at 
fixed halo mass \citep{Yang03,Bosch03a,Bosch03c}. Our predictions can be tested
by analyses of large galaxy redshift surveys like the Two Degree Field
Galaxy Redshift Survey (2dFGRS; \citealt{Colless01}) and the 
Sloan Digital Sky Survey (SDSS; \citealt{York00}).  Several of our
qualitative predictions are in good agreement with recent analyses
of these surveys, in particular with \citeauthor{Zehavi05}'s 
(\citeyear{Zehavi05}) study of the luminosity and color dependence of the 
SDSS galaxy correlation function, as we discuss in \S\ref{sec:disc}.

\section{Theoretical Models}
\label{sec:models}

We use the same SPH simulation and SA galaxy formation model as B03.
We review these calculations briefly here and refer the reader to B03
for more details.

The SPH simulation assumes a cosmological model with $\Omega_m=0.4$, 
$\Omega_\Lambda=0.6$, $\Omega_b=0.02 h^{-2}$, $h\equiv H_0/(100\,{\rm km s^{-1}
Mpc^{-1}})=0.65$, $n=0.95$, and $\sigma_8=0.8$ (see descriptions by 
\citealt{Murali02,Dave02,Weinberg04}). It follows the evolution
of $144^3$ gas and $144^3$ dark matter particles in a box of $50 \hMpc$ on 
each side from $z=49$ to $z=0$. The softening length of the gravitational 
force is $7 h^{-1} {\rm kpc}$ comoving. 
The simulation incorporates radiative and Compton cooling 
and phenomenological prescriptions for star formation and supernova
feedback. Galaxies are identified as gravitationally bound groups of star and 
cold gas particles that are associated with a common local maximum
in the baryon density.  Dark matter
halos are identified using a friends-of-friends algorithm (\citealt{Davis85})
with a linking 
length of 0.173 times the mean interparticle separation. Each galaxy is 
assigned to the halo that contains the dark matter particle closest to the
galaxy center of mass. In each halo, the galaxy whose center is closest
to the position of the most bound dark matter particle (defined as the halo 
``center") is tagged as the ``central" galaxy, while others are regarded as
satellite galaxies. 

The SA model used by B03 is GALFORM (\citealt{Cole00}).  For a given halo, 
the model generates a ``merger tree" using a Monte Carlo method, 
starting at $z=0$ and branching into progenitor halos until it reaches 
a starting redshift. Then the model works forward in time to follow the 
formation and evolution of galaxies in each progenitor halo. Phenomenological
prescriptions are used to model star formation and feedback, dynamical 
friction within halos, and mergers of galaxies. There is always a galaxy 
residing at the center of each halo. Before a halo experiences a major 
merger, cooling gas is assumed to accrete onto the disk of this galaxy
and form stars. If two halos merge, the most massive galaxy is set to 
become the central galaxy of the merged halo, and any other galaxies 
become satellites. If two galaxies of comparable mass merge, then all their 
stars (disk+spheroid) form the spheroid of the remnant, which may regrow a
new disk by subsequent gas accretion. Some adjustable parameters of 
this model are chosen on the basis of observed properties of the local 
galaxy population, most notably the galaxy luminosity function.
However, no parameters are adjusted to reproduce observed galaxy 
clustering or the SPH results.  The model uses the same cosmological 
parameters as the SPH simulation, and it is supplied with the same halo 
population identified in the SPH simulation so that the HOD predictions 
of these two models can be compared halo by halo.  To improve statistics, 
ten SA realizations are conducted for each of the 70 most massive SPH halos, 
and forty realizations are done for less massive halos in each mass bin of 
width $\Delta \log M=0.1$.
   
\section{Halo Occupation Distributions}
\label{sec:hod}
 
Galaxy samples of different space densities are constructed by choosing
galaxies above different baryonic mass thresholds. 
The baryonic mass resolution limit of the SPH simulation (corresponding to 64
SPH particles) is $5.42 \times 10^{10}M_\odot$, and the space density
of simulated galaxies above this threshold is 
$\ngavg=0.02 h^3 {\rm Mpc^{-3}}$.  The baryonic mass threshold that
yields the same space density in the SA model is 
$1.45 \times 10^{10} \Msun$, lower because of the suppressed
gas cooling and enhanced supernova feedback (see B03 for further discussion).
In \S\ref{sec:hod_all} we focus on this sample with 
$\ngavg=0.02 h^3 {\rm Mpc^{-3}}$, which is the largest one we
can create from the SPH simulation.  It should correspond roughly to 
an observational sample defined by a luminosity threshold
$M_r=-18.6$ ($0.18 L_*$), where we have used the \cite{Blanton03}
luminosity function to find the luminosity threshold that yields the
same comoving space density.  Scatter in stellar mass-to-light ratios
makes mass-threshold and luminosity-threshold samples different,
but the substantial variations of these mass-to-light ratios across
the galaxy population are largely a systematic function of 
luminosity \citep{Kauffmann03}, with only moderate scatter at fixed $L$.

In \S\ref{sec:hod_age}, we divide this sample into two equal parts
on the basis of stellar population age --- the median lookback time
(when half the stars had formed) in the case of the SPH simulation
and the mean lookback time in the case of the SA model. 
The SA code automatically computes the mass-weighted mean stellar
age, but it does not produce the step-by-step stellar mass record needed
to compute the median stellar age {\it a posteriori}.  The SPH code
does produce this record, making the median age easier to compute;
this quantity is probably more robust than the mean age in the SPH 
simulations because star formation rates are underestimated in a 
galaxy's early history, when it is near the mass resolution threshold.
Our analysis uses only the
rank order of ages not the absolute values, so we expect the
difference in age definitions to have minimal impact on our results.
Population age should be a fairly good proxy for galaxy color 
or spectral type, so this division mimics the red/blue or early/late
divisions studied by \citeauthor{Zehavi02} (\citeyear{Zehavi02};
\citeyear{Zehavi05}) in the SDSS and \cite{Norberg02} and 
\cite{Madgwick03} in the 2dFGRS.  In \S\ref{sec:hod_mass}, we consider
samples of progressively higher baryonic mass threshold and lower
mean space density, to investigate the predicted luminosity dependence
of the HOD.  Although we present both SPH and SA results throughout
this section, the SA model has better statistics because of the
multiple realizations of each halo, and it thus allows us to investigate
some finer points of the HOD predictions.

\subsection{HOD for All Galaxies}
\label{sec:hod_all}

\begin{figure}
\plottwo{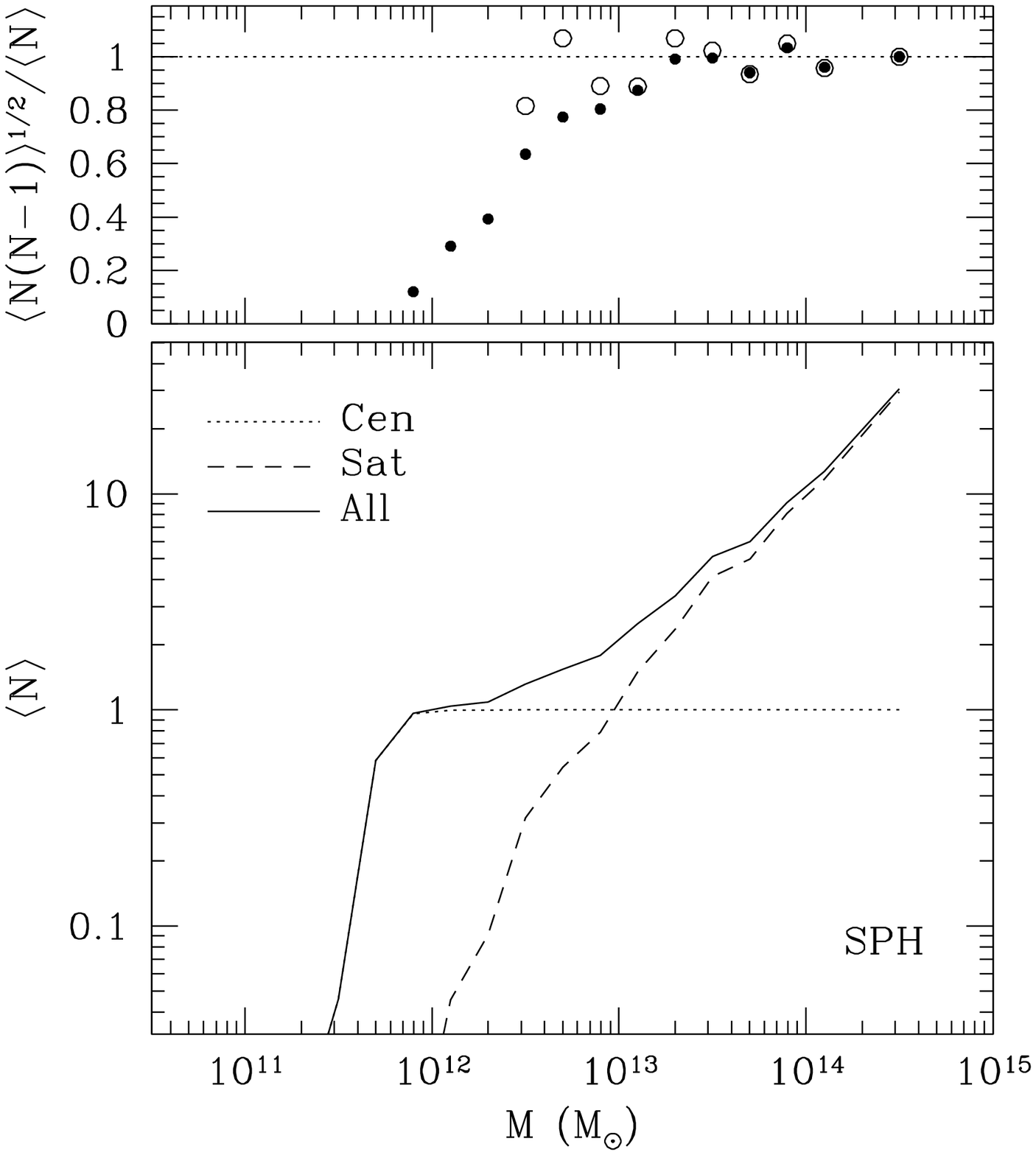}{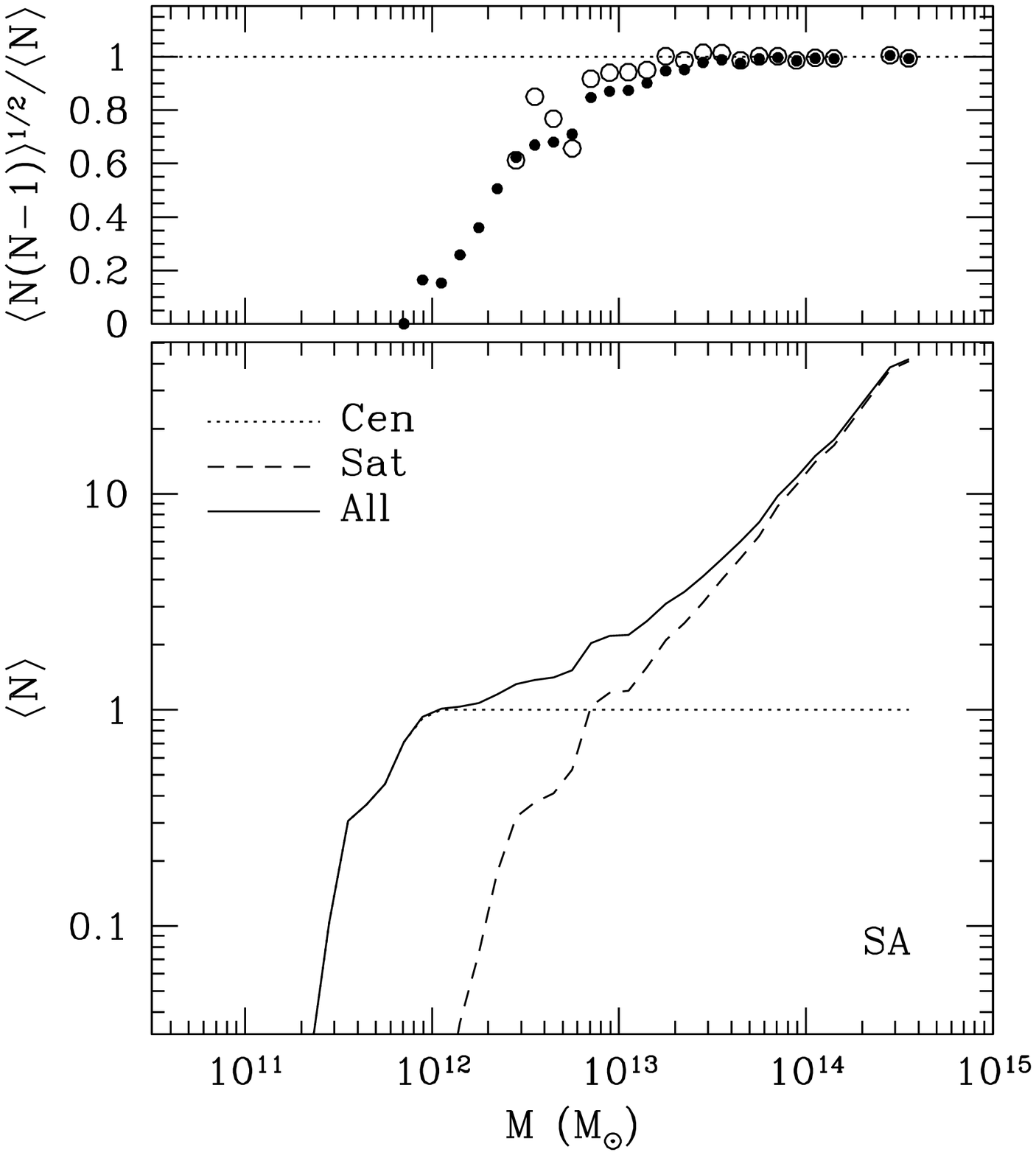} 
\caption[]{\label{fig:hod}
Mean occupation number and scatter as a function of halo mass, separated 
into central and satellite galaxies. Predictions are shown for the 
$\ngavg=0.02 h^3 {\rm Mpc}^{-3}$ samples from the SPH simulation (left panels)
and from the SA model (right panels). Lower panels plot the mean occupation 
numbers of central, satellite, and all galaxies. In the upper panels, circles
show $\langle N(N-1)\rangle^{1/2}/\langle N\rangle$, indicating the width of 
the probability distribution, for all galaxies (filled circles) and satellite
galaxies (open circles). For Poisson $P(N|M)$, this ratio would be one (dotted 
line). This figure can be compared to Fig.~4 of K04. 
}
\end{figure}

Figure~\ref{fig:hod} shows mean occupation numbers as a function of 
halo mass predicted by the SPH simulation and by the SA model, 
for the $\ngavg = 0.02 h^3 {\rm Mpc}^{-3}$ samples. Mean occupation 
functions for central and satellite galaxies are similar to those
found for subhalos by K04 (see their Figure~4). The total mean occupation
function is the sum of a step-like function representing the contribution 
of central galaxies and a power-law-like function representing the 
contribution of satellite galaxies. 

If each sample is constructed by first selecting dark matter halos above a 
minimum mass, the mean occupation function $\Ncenavg$
of central ``subhalos'' (really the halos themselves) 
would be a strict step function, since $\Ncen=0$ below the minimum
mass and $\Ncen=1$ above it.  However, our samples are based on minimum 
galaxy {\it baryonic} masses, so scatter in the relation between 
baryonic mass of the central galaxy 
and virial mass of the halo smooths the step, with 
$\Ncenavg$ increasing from 0.1 to 0.9 over a factor $\sim 2$--3 in 
halo mass. Since a halo necessarily contains either
zero or one central galaxies, the 
probability distribution $P(\Ncen|\Ncenavg)$ is a
nearest-integer distribution (more technically, a Bernoulli distribution),
with $P(1) = 1-P(0) = \Ncenavg$.

The mean occupation function of satellite galaxies in both the SPH and SA
calculations is roughly a power law, $\Nsatavg \propto M^\alpha$, though
it tails off more rapidly at masses where $\Nsatavg < 1$.  In fact, the
power law index is $\alpha \approx 1$ in both cases, implying
a simple proportionality between halo mass and satellite number.
The mean number of galaxy pairs in a halo, $\langle N(N-1) \rangle$,
is important for the small scale behavior of the galaxy two-point correlation
function.  Upper panels of Figure~\ref{fig:hod} plot the quantity
$\langle N(N-1) \rangle^{1/2}/\langle N \rangle$, as a function of halo
mass. If $N$ is Poisson distributed, then this quantity is unity. 
Satellite galaxies have
$\langle N(N-1) \rangle^{1/2}/\langle N \rangle \approx 1$ for all masses
where $\Nsatavg > 1$ (open circles). In the regime where $\Nsatavg\sim 1$,
the SA model seems to predict a value of 
$\langle N(N-1) \rangle^{1/2}/\langle N \rangle$ slightly lower than unity. 
The width of the total galaxy
$P(N|M)$ (filled circles) is close to Poisson at high masses, where satellite
galaxies dominate the occupation number, but it is substantially sub-Poisson
($\langle N(N-1) \rangle^{1/2}/\langle N \rangle < 1$) at low masses
because the nearest-integer distribution for central galaxies is narrower,
with $\Ncen(\Ncen-1) \equiv 0$.  
Our findings that satellite galaxies have a power-law-like $\Nsatavg$ with
$\alpha \approx 1$, that satellite numbers follow a roughly Poisson 
distribution with respect to their mean, and that the sub-Poisson fluctuations
of the total galaxy $P(N|M)$ are thus a consequence of central galaxies,
are all in agreement with the findings of K04 for dark matter subhalos.

\begin{figure}[h]
\plottwo{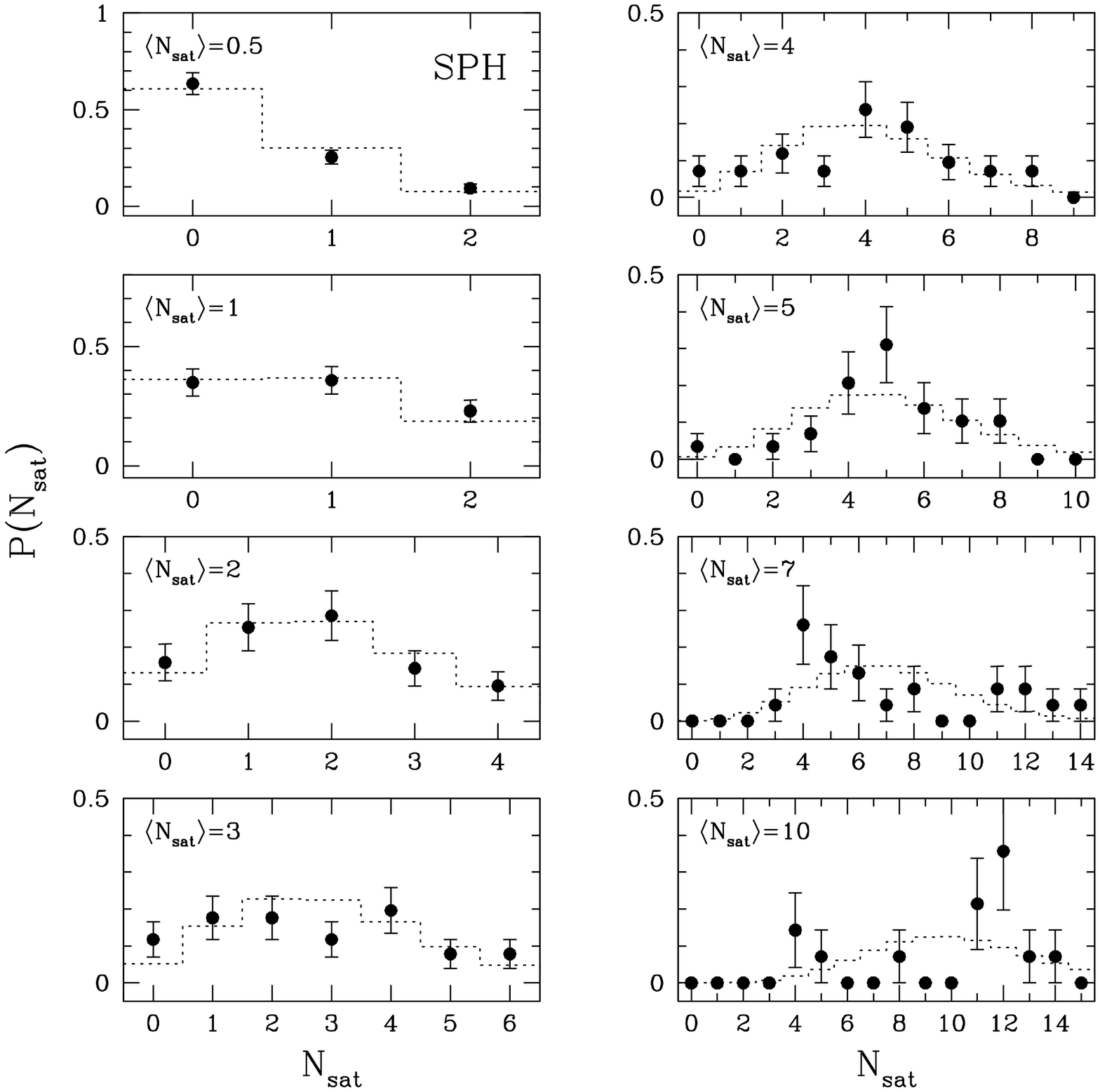}{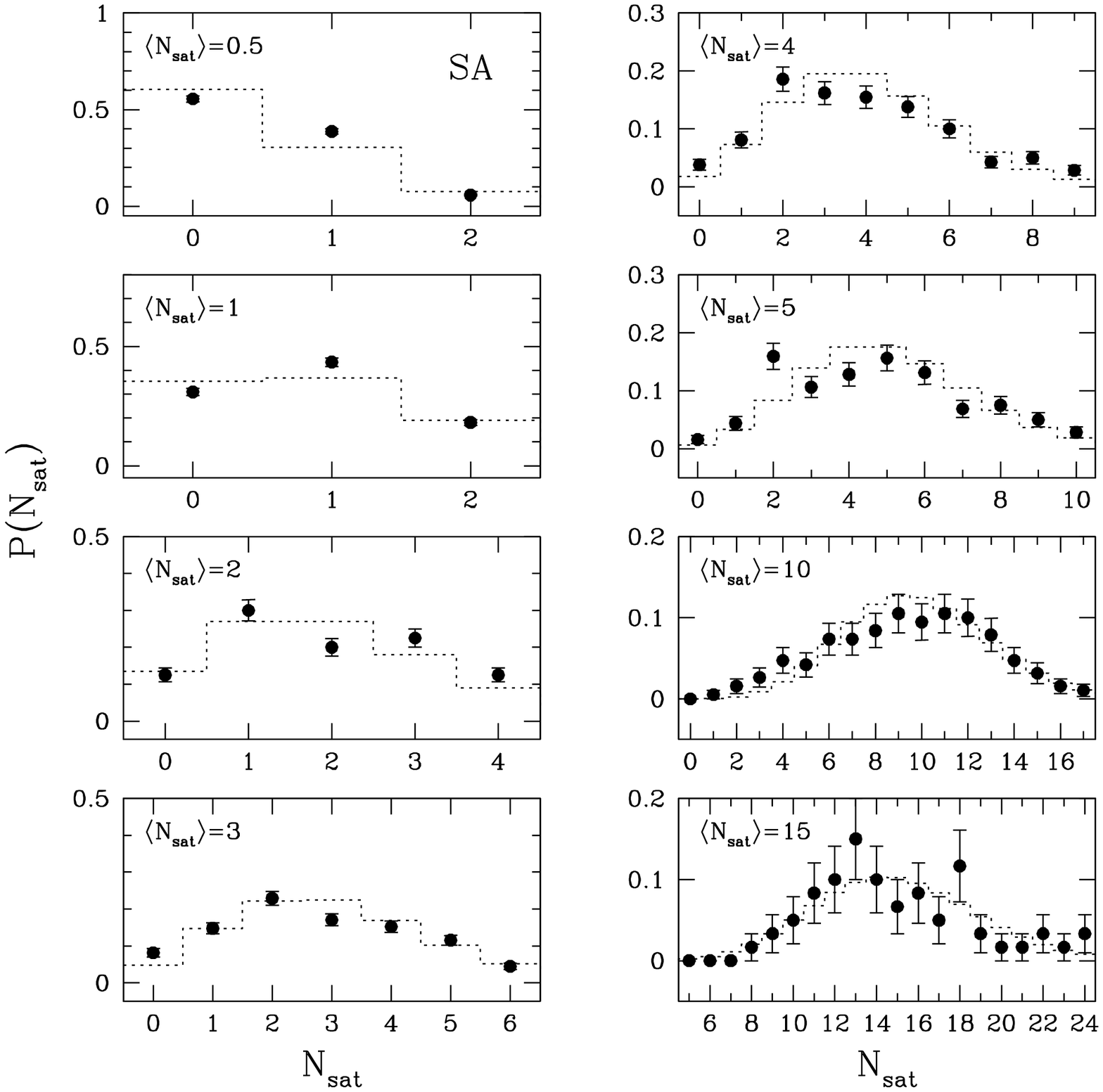}
\caption[]{\label{fig:PNM}
Probability distributions of satellite numbers as a function of 
the mean occupation number of satellites, predicted by the SPH simulation 
(left panels) and by the SA model (right panels). Points are predictions of 
the models, and the Poisson error in each bin is assigned as the error bar. 
The dotted histogram in each panel shows the Poisson distribution of the 
same mean.
}
\end{figure}

To test how closely the probability distribution of satellite occupation 
number follows a Poisson distribution, 
Figure~\ref{fig:PNM} plots $P(\Nsat|\Nsatavg)$ at different values of 
$\Nsatavg$ (points), in comparison to a Poisson distribution that has the 
same mean (dotted histogram). For both the SPH and the SA models, the typical 
bin width of halo mass is  $\Delta\log M$=0.4, but it is set to be 0.1 for 
the SA model if $\Nsatavg\leq 2$.
As a result of multiple realizations, the SA model gives better statistics 
than the SPH simulation, especially for 
high occupation numbers that correspond to rare, massive halos. Nevertheless, 
both models predict that $P(\Nsat|\Nsatavg)$ is impressively close to a Poisson 
distribution over the full range $\Nsatavg=0.5$ to $\Nsatavg=15$ where we
have adequate statistics.  While visual inspection of Figure~\ref{fig:PNM}
suggests that the match is not exact (we have not carried out a formal
statistical test), it shows that the Poisson approximation for 
$P(\Nsat|\Nsatavg)$ is likely to be adequate for most predictions of
galaxy clustering statistics, such as counts-in-cells distributions
and higher order correlation functions.

To model two-point correlation functions of SDSS galaxies, 
\citet{Zehavi05} parameterize the mean occupation function for galaxies 
brighter than a luminosity threshold as a step function for $\Ncenavg$ plus 
a truncated power law $\Nsatavg=(M/M_1)^\alpha$ for satellites. 
This simple model has three free parameters: the minimum halo mass $\Mmin$ 
below which $\Ncen=\Nsat \equiv 0$ and the slope $\alpha$ and normalization
$M_1$ of the power law.  For a given cosmology, two parameters can be adjusted
to fit the observed correlation function while the third (usually $\Mmin$)
is set by matching the mean galaxy
number density of the sample. 
Figure~\ref{fig:hod} shows that this parameterization captures the
main features of the theoretically predicted HOD.  It has the benefit
of having the same number of free parameters as a power law (since one
parameter is fixed by the galaxy number density), allowing a fair
comparison of goodness-of-fit to observed correlation functions.
However, as measurements of complementary clustering statistics
become available, we can afford to fit HOD models with more free
parameters, and they may even become necessary to match the data.
For example, with the high mass end of $\Navg$ constrained by the
group multiplicity function (e.g., 
\citealt{Peacock00,Marinoni02,Kochanek03,Yang05b};
A. Berlind, et al., in preparation), we can use the correlation function 
and other clustering statistics to 
investigate the cutoff profile of the mean occupation function. 
Therefore, it is useful to more accurately parameterize the results from 
the theoretical models with slightly more complicated prescriptions. 
Ultimately, we would like to fit observations
using a model that does not rely on a theoretically predicted form, 
but in the near term we can use the predicted form and compare observationally
inferred and theoretically predicted parameter values.
It is worth noting that if the adopted cosmological model is substantially
wrong then {\it no} choice of HOD can match the full range of 
galaxy clustering statistics (Z. Zheng \& D. Weinberg, in preparation; 
see \citeauthor{Bosch03c} [\citeyear{Bosch03c}] for closely related arguments 
using the conditional luminosity function).

For both the SPH and SA models, we find that the mean occupation function 
for central galaxies can be well represented by 
\begin{equation}
\label{eqn:Ncenerf}
\Ncenavg = \frac{1}{2}\left[1+{\rm erf}
  \left(\frac{\log M-\log\Mmin}{\sigM}\right)\right], 
\end{equation}
where {\tt erf()} is the error function   
\begin{equation}
{\rm erf}(x) = \frac{2}{\sqrt\pi} \int_0^x e^{-t^2} dt.
\end{equation}
The parameters are $\Mmin$, the characteristic minimum mass of halos that 
can host such central galaxies, and $\sigM$, the characteristic 
transition width. 
This functional form corresponds to a Gaussian distribution of
$\log \Mgal$, at fixed halo mass $M$. If the 
mean baryonic mass is $\langle \Mgal \rangle \propto M^\mu$
near $\Mmin$, then $\sigM=\sigMg/\mu,$ where $\sigMg$ is the
scatter in the logarithm of galaxy baryonic mass at fixed halo mass.  
For a sample defined
by a luminosity threshold instead of a baryonic mass threshold, the halo mass
dispersion $\sigM$ will be somewhat larger because of the scatter in
stellar mass-to-light ratios, and the scatter may deviate from the Gaussian 
distribution.  The observed Tully-Fisher (\citeyear{Tully77}) relation 
(roughly $\langle L \rangle \propto M$, $\sigma_{\log L} \sim 0.15$) suggests 
$\sigM \sim 0.15$ in the mass range $M \sim 10^{12}M_\odot$ corresponding to 
typical bright spirals.  (All logarithms in this paper are base 10.)
A detailed discussion of the scatter in the 
mass-luminosity relation and its effects can be found in 
\citet{Tasitsiomi04}.

\begin{figure}[h]
\plottwo{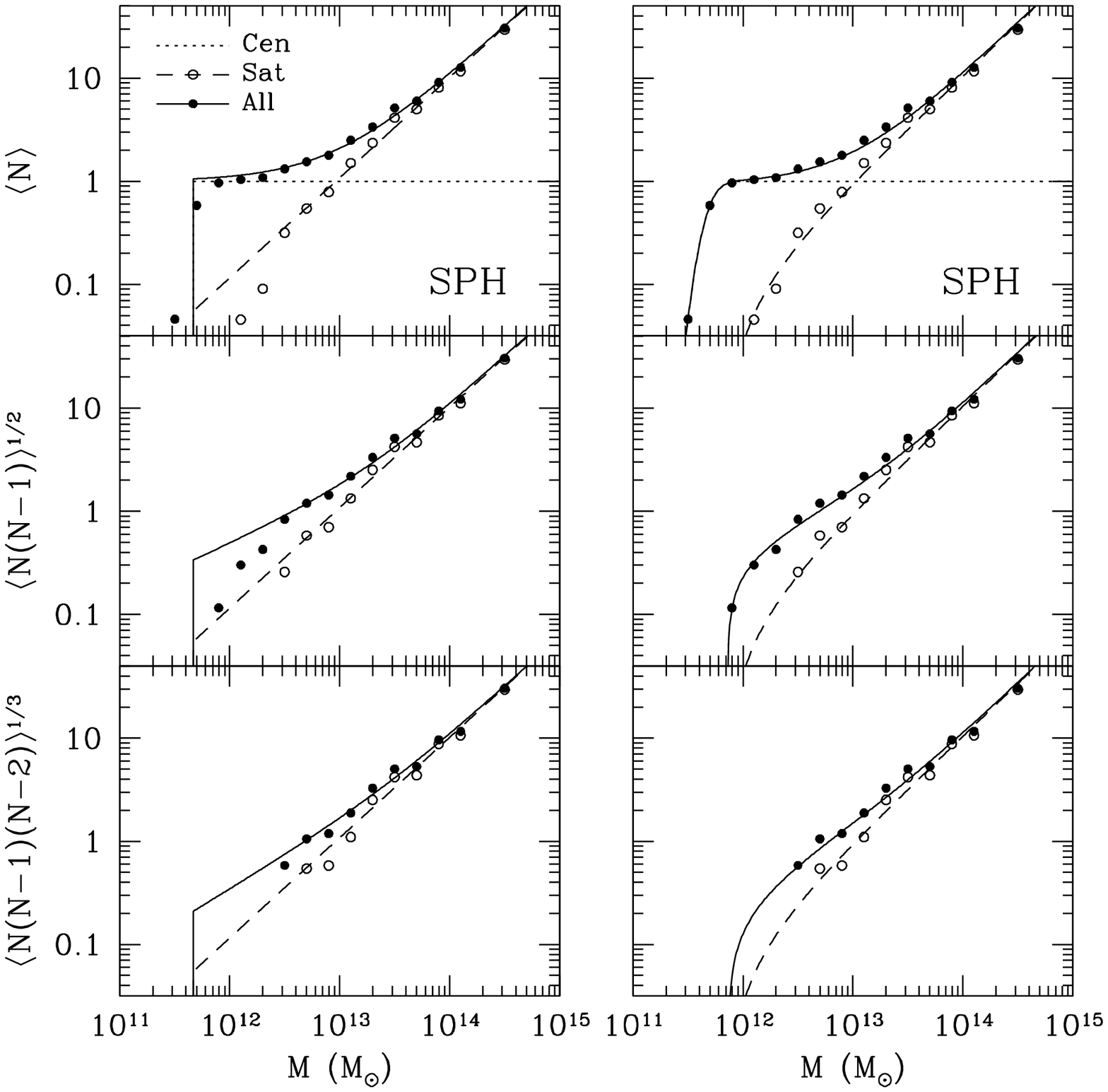}{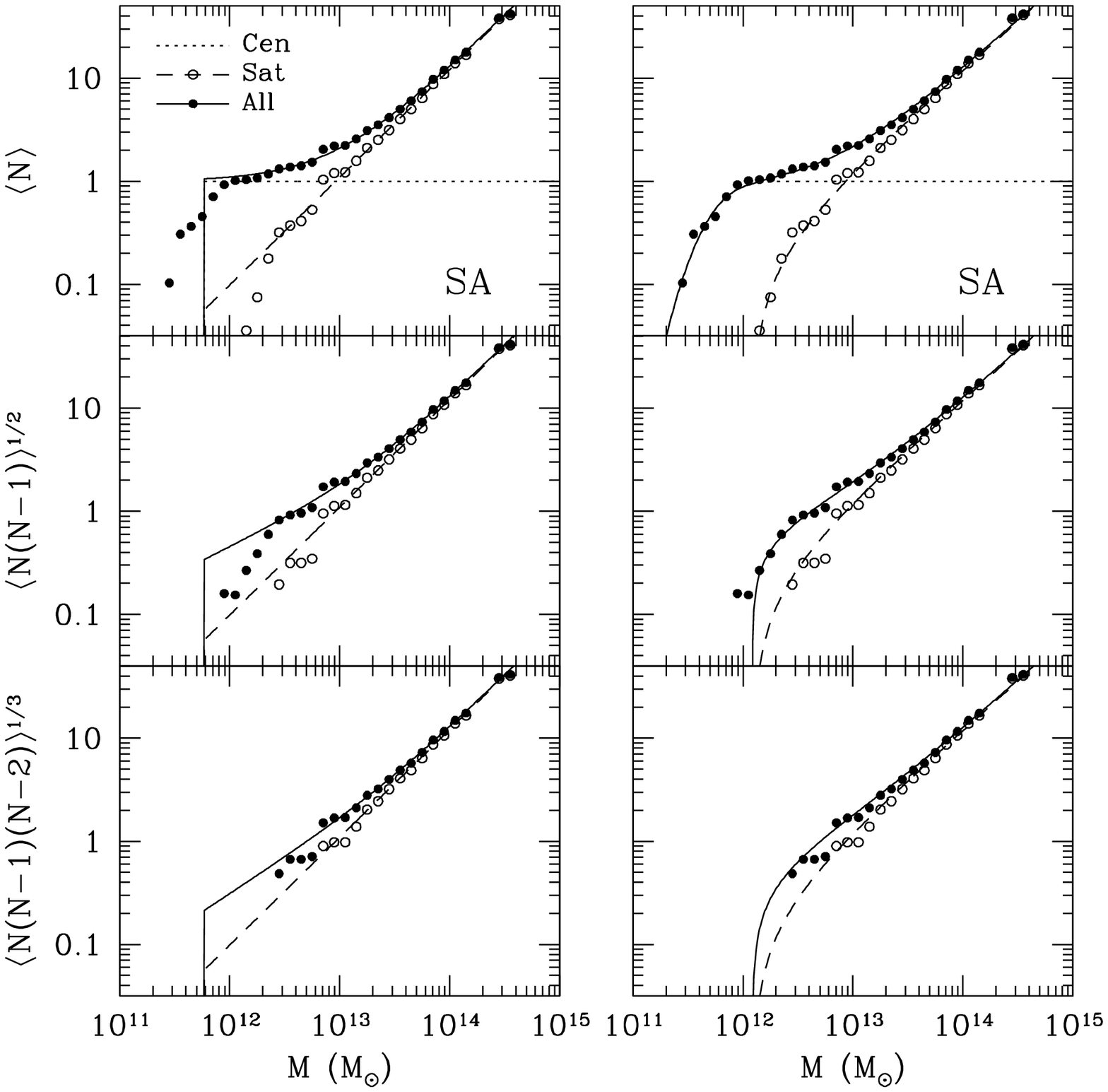}
\caption[]{\label{fig:parfit}
Parameterized fits to mean occupation functions (top panels) and predicted 
numbers of galaxy pairs and triplets (middle and bottom panels) for the 
SPH simulation (left)
and the SA model (right). For each model, left panels show 
results based on 3-parameter fits, which assume sharp cutoff profiles of 
$\Ncenavg$ and $\Nsatavg$, and right panels show results of fits with more 
parameters to model the cutoff profiles (see eqs.~[\ref{eqn:Ncenerf}] and 
[\ref{eqn:Nsat}]). 
Fits and predictions are plotted as curves, and circles are 
measurements from the models. 
}
\end{figure}

At low masses, 
the mean occupation function for satellite galaxies drops below a
power law extrapolation of $\Nsatavg$ from high masses.
Similar to K04, we find that the full range of $\Nsatavg$ can be well 
approximated by the form
\begin{equation}
\label{eqn:Nsat}
\Nsatavg = [(M-M_0)/M^\prime_1]^\alpha
\end{equation}
for $M>M_0$, where the truncation mass $M_0$ for satellites may
differ from the truncation mass $\Mmin$ for central galaxies.
Note that with this parameterization, $M^\prime_1$ is {\it not} the mass
at which $\Nsatavg = 1$.   The top panels of Figure~\ref{fig:parfit}
show fits of the 3-parameter ($\Mmin$, $M_1$, $\alpha$) and 5-parameter
($\Mmin$, $\sigM$, $M^\prime_1$, $M_0$, $\alpha$) models to the SPH and SA
mean occupation functions.  The parameter values are listed in 
Table~\ref{tbl:params}.  
Note that these values are likely to depend on the cosmological model
as well as the galaxy space density and the physics encoded in the SPH
and SA calculations.  The 3-parameter form (used by \citealt{Zehavi05})
captures the results of both models well but not perfectly, while
the 5-parameter form gives a nearly perfect fit in both cases.

If we assume that central and satellite galaxies 
have nearest-integer and Poisson distributions, respectively, then 
we can calculate all the higher 
moments of the occupation number based on the fits to the mean occupation.
In particular, we can predict average numbers of galaxy pairs and 
triplets inside halos, which are relevant to the 1-halo terms of the
galaxy 2-point and 3-point correlation functions, respectively.
The total occupation number $N$ is the sum of 
$\Ncen$ and $\Nsat$, and it is easy to show that 
\begin{equation}
\label{eqn:NNm1}
\langle N(N-1) \rangle =   \langle \Ncen(\Ncen-1) \rangle 
                         + 2 \langle \Ncen \Nsat \rangle
                         + \langle \Nsat(\Nsat-1) \rangle
\end{equation} 
and
\begin{equation}
\label{eqn:NNm1Nm2}
 \begin{array}{ll}
 \langle N(N-1)(N-2) \rangle = &   \langle \Ncen(\Ncen-1)(\Ncen-2) \rangle 
                                  + 3 \langle \Ncen(\Ncen-1)\Nsat \rangle \\
                               &  + 3 \langle \Ncen\Nsat(\Nsat-1) \rangle 
                                  + \langle \Nsat(\Nsat-1)(\Nsat-2) \rangle.
 \end{array}
\end{equation}
Since $\Ncen=0$ or 1, the first term on the right hand 
side (RHS) of equation~(\ref{eqn:NNm1}) and the first two terms on the RHS 
of equation~(\ref{eqn:NNm1Nm2}) vanish. 
The first surviving terms represent pairs or triplets involving the
central galaxy, while the last terms represents combinations that only
involve satellites.  To deal with the cross-correlation 
between occupations of central and satellite galaxies, we simply
need to note that if $\Ncen=0$ then $\Nsat=0$ by definition.
Therefore, $\langle \Ncen \Nsat \rangle = 
\langle \Nsat \rangle$ and $\langle \Ncen \Nsat(\Nsat-1) \rangle = 
\langle \Nsat(\Nsat-1) \rangle $.  Assuming a Poisson distribution for
$\Nsat$,
equations~(\ref{eqn:NNm1}) and (\ref{eqn:NNm1Nm2}) reduce to
\begin{equation}
\label{eqn:NNm1_f}
\langle N(N-1) \rangle = 2 \langle\Nsat\rangle
                         + \langle\Nsat\rangle^2 
\end{equation}
and
\begin{equation}
\label{eqn:NNm1Nm2_f}
\langle N(N-1)(N-2) \rangle = 3 \langle\Nsat\rangle^2 
                              + \langle\Nsat\rangle^3.
\end{equation}
We plot predictions based on these two equations and the fits to $\Ncenavg$ 
and $\Nsatavg$ in the middle and bottom panels of Figure~\ref{fig:parfit}. 
Points in these panels are measurements from the SPH and 
SA calculations.
We see that the 3-parameter model overpredicts the number of galaxy 
pairs and triplets in low mass halos, 
mainly because the number of satellites is overpredicted by ignoring the 
profile of the low mass cutoff.
For the 5-parameter fit to $\Navg$, the predicted numbers
of pairs and triplets agree remarkably well with the measured values, 
providing further evidence that the Poisson approximation for 
$P(\Nsat|\Nsatavg)$ is adequate for practical calculations.

 \subsection{HOD for Young and Old Galaxies}
 \label{sec:hod_age}

The top
panels of Figure~\ref{fig:color_hod} show mean occupation functions
for the young and old halves of the $\ngavg=0.02 h^3 {\rm Mpc}^{-3}$
sample, in comparison to $\Navg$ of the full sample.
Solid curves in the bottom panels show
the fraction of young galaxies as a function of halo mass. These are 
basically the same as in Figure~13 of B03. While the mean 
occupation number of old galaxies rises continuously as halo mass increases,
the $\Navg$ of young galaxies first rises, then declines to a local minimum
at $M \sim 10^{13} \hMsun$, then rises again.  \citet{Sheth01} find 
similar non-monotonic behavior for the mean occupation of blue galaxies in
the SA models of \citet{Kauffmann99}. The fraction of young galaxies
(lower panel) has a steep drop at low occupation number and decreases
slowly toward higher occupation number. 

The shapes of these mean occupation functions are easy to understand
if we separate contributions from central and satellite galaxies,
as shown in the middle panels of Figure~\ref{fig:color_hod}. 
As shown by B03 (their Fig.~19), both the SPH simulation and the SA model 
predict that on average the (median/mean) stellar age of a halo's central 
galaxy is an increasing function of halo mass. 
As explained by B03, this correlation between stellar age of
the central galaxy and the mass of the parent halo arises from two physical
effects: higher mass halos begin to assemble earlier, allowing an
earlier onset of star formation in the central galaxy, and gas accretion
rates drop once a halo becomes massive enough to support a virial
shock \citep{Keres04}, choking off the formation of young stars at late times.
In halos of mass near $\Mmin$, therefore, central galaxies are 
usually young (i.e., below the sample's median age), while in high mass
halos they are all old.  The minimum in $\Navg$ of young galaxies
occurs for halos that are too massive to host young central galaxies
but not massive enough to host satellite galaxies. 
The typical age of satellite galaxies also increases with halo mass,
but the correlation is weaker (see B03, Fig.~19) because satellites 
experience most of their growth in lower mass halos that merge
into their final host halo.  In Figure~\ref{fig:color_hod}, the
young galaxy fraction for central galaxies drops rapidly with 
increasing halo mass and falls essentially to zero, while the
young galaxy fraction for satellites drops more slowly and
in the SA calculation reaches a nearly flat plateau at high mass.
The different HODs of young and old galaxies reflect fundamental
aspects of galaxy formation physics, and their origin is
transparent once we separate central and satellite contributions.

As with the full sample HODs, we would like to find simple parameterized
prescriptions that capture these age-dependent features, which we can 
do by describing the young galaxy central and satellite fractions as
a function of halo mass.
Based on the bottom panels of Figure~\ref{fig:color_hod}, we adopt
\begin{equation}
f_{\rm young,cen} = f_c  
  \left[1+\exp\left(\frac{\log M-\log M_c}{\sigMc}\right)\right]^{-1},
\end{equation} 
where $f_c$ sets the amplitude and $M_c$ and $\sigMc$ characterize the 
transition mass scale and speed. 
We approximate the fraction of young satellite galaxies by an exponential,
\begin{equation}
f_{\rm young,sat} = f_s \exp\left({-\frac{\log M-\log M_0}{\sigMs}}\right),
\end{equation}
where $M_0$ is the same satellite cutoff mass used in equation~(\ref{eqn:Nsat}) 
and $\sigMs$ characterizes the speed of the falloff.
Top panels of Figure~\ref{fig:parfit_c} show that the above functional forms
for the $\ngavg=0.02 h^3 {\rm Mpc^{-3}}$ samples,
together with our 5-parameter model for the full galaxy $\Navg$,
allow accurate fits to the SPH and SA mean occupation functions.
The parameters we use for the SPH(SA) model are $f_c=0.71(0.65)$, 
$\log M_c=12.55(12.64)$, $\sigMc=0.26(0.14)$, $f_s=0.99(0.80)$, and 
$\sigMs=1.50(1.10)$.

To predict numbers of galaxy pairs and triplets, we need additional 
assumptions. We cannot automatically simplify the central-satellite
cross terms of equations~(\ref{eqn:NNm1}) and~(\ref{eqn:NNm1Nm2})
to obtain equations~(\ref{eqn:NNm1_f}) and~(\ref{eqn:NNm1Nm2_f})
because a halo may, for example, have an old central galaxy
($N_{\rm cen,young}=0$) but nonetheless have young satellites.
To proceed, we assume that the satellite population in a halo
of a given mass does not depend on the age of the central galaxy, 
and that the occupation numbers of young and old satellite galaxies
follow independent Poisson distributions with respect to their
individual means.  Equations ~(\ref{eqn:NNm1_f}) and~(\ref{eqn:NNm1Nm2_f}) 
are then modified by changing the first RHS terms to
$2 \langle\Ncen\rangle \langle\Nsat\rangle$ and 
$3 \langle\Ncen\rangle \langle\Nsat\rangle^2$, respectively. The middle and 
bottom panels of Figure~\ref{fig:parfit_c}, which compare pair and triplet
predictions from the mean occupation fits to the 
SPH and SA results, suggest that the
above approximations are accurate enough for calculations of galaxy clustering.

Figure~\ref{fig:Mgalbin} shows HODs for samples defined by a bin of baryonic
mass instead of a threshold; specifically, we take the less massive half
of our $\ngavg=0.02h^3{\rm Mpc}^{-3}$ samples, and we again divide into
young and old subsamples.  The shapes of the mean occupation functions
for all, young, and old galaxies look alike --- a bump at lower halo masses
resulting from central galaxies and an approximately power-law function toward 
higher halo masses representing the satellite contribution.
\citet{Guzik02} find a similar result for a luminosity-bin sample using
the \citet{Kauffmann99} SA model.  For our chosen mass bin, the majority 
of central galaxies are young, and (since the young and old samples are roughly
equal) the majority of satellites are therefore old. Lower mass bins have
a higher fraction of young central galaxies, while high mass bins have
predominantly old central galaxies. The 3-parameter and 5-parameter models 
described in \S~\ref{sec:hod_all} are not appropriate for describing the HOD 
of a mass or luminosity bin because they do not allow an upper cutoff in 
$\Ncenavg$. However, these models do provide good descriptions for samples 
defined by mass or luminosity thresholds, and the HOD of a bin sample is
just the difference of the HODs for thresholds at the bin's upper and lower
limits by mass or luminosity thresholds, and the HOD of a bin sample is
just the difference of the HODs for thresholds at the bin's upper and lower
limits.

 \subsection{HOD Parameters as a Function of Galaxy Mass/Luminosity}
 \label{sec:hod_mass}

As the threshold baryonic mass is increased, the predicted mean occupation
function maintains the same general form but shifts horizontally towards
higher masses in the $\log\Navg$--$\log M$ plane (B03, Fig.~9).
Figure~\ref{fig:MminM1} examines the dependence of characteristic
halo mass scales for central and satellite galaxies on the baryonic
mass threshold.  The lower solid curves represent $\Mmin$, the characteristic 
minimum mass for hosting a central galaxy above the threshold.
Since $\Ncenavg$ is not a strict step function, we simply define $\Mmin$ 
in this figure as the mass at which $\Navg=0.5$; half of the halos with 
$M=\Mmin$ contain a central galaxy above the threshold, and half do not.
This halo mass threshold is linearly proportional to the baryonic
mass threshold, $\Mmin \propto \Mgal$, for $\Mmin \la 3\times 10^{12}\hMsun$,
implying that the baryonic mass of the central galaxy is 
proportional to the mass of the host halo in this regime.
The dot-dashed line in each panel marks $\Mmin = (\Omega_b/\Omega_m)\Mgal$,
expected if the ratio of central galaxy mass to halo virial mass is
equal to the universal baryon fraction.  The SPH results lie close
to this limiting case for low halo masses, while the SA central galaxies
never accrete more than about 25\% of the available baryons.
As discussed by B03, this difference can be attributed to treatments of the 
gas core radius and stellar feedback in the SA model, which are adjusted to 
fit the observed galaxy luminosity function.  The SPH simulation predicts 
excessively massive galaxies, assuming a standard stellar initial mass 
function (IMF).  Additional processes such as photoionization by the UV 
background or supernova gas blowout may suppress accretion in very low mass 
halos and produce an upturn in the $\Mmin$--$\Mgal$ relation, but they are 
not evident in the mass range probed here.

In high mass halos, some of the baryonic mass goes into satellite galaxies,
and the efficiency of gas cooling drops so that a smaller fraction 
of baryons are accreted onto galaxies (B03, Fig.~5).  As a result, $\Mmin$ 
begins to grow faster than $\Mgal$ at masses 
$\Mmin \ga 3\times 10^{12}\hMsun$, corresponding to group or cluster halos.
The SA model predicts a much steeper rise than the SPH simulation,
a consequence of resolution effects in the SPH simulation and of the different
ways the two models treat cooling and feedback.  The SPH simulation includes 
thermal feedback from supernovae, but the thermal energy is usually deposited 
in dense gas and radiated away before it can drive a galactic wind 
(\citealt{Katz96}).  This relatively efficient cooling produces overly 
luminous galaxies at the centers of groups and clusters (again assuming a 
standard stellar IMF), though the baryonic masses and hence luminosities are
reduced somewhat when the numerical resolution is increased
(see B03, Fig.~5).  In the SA model, the formation of very bright 
galaxies is suppressed by adjusting the core radius in the gas density 
profile, which controls how much of the gas can cool, so that the model 
matches the bright end of the observed galaxy luminosity function 
(\citealt{Cole00}).  The core radius mechanism is somewhat artificial, and 
the ``solution''
in the real universe probably involves some additional physics such
as thermal conduction or AGN-driven superwinds (see \citealt{Benson03a})
or reduced cooling efficiency in multi-phase halo gas \citep{Maller04}.
However, the trend of $\Mmin$ with $\Mgal$ is likely to resemble
the SA curve shown here for any model that matches
the observed galaxy luminosity function, since the halo mass function
is fixed by the cosmology and the luminosity function by observations.

The upper solid curves in each panel of Figure~\ref{fig:MminM1} show the 
dependence of $M_1$ on galaxy baryonic mass, where $M_1$ is the mass of 
halos that on average contain one satellite galaxy (i.e., $\Navg=2$)
above the baryonic mass threshold.  These curves look remarkably
like scaled versions of the $\Mmin$ curves.  The dashed curves
represent 14$\Mmin$ and 18$\Mmin$ for the SPH and SA models, respectively,
and they match the $M_1$ curves almost perfectly 
except for very massive halos.  The smaller scaling factor in the
SPH simulations is at least partly a consequence of the more
efficient cooling onto central galaxies, discussed above, which
leads the simulation to produce
galaxies that are increasingly overluminous as halo mass increases. 
The large gap between $\Mmin$ and $M_1$, more than an order of magnitude
in both calculations, produces the plateau in the mean occupation
function between $\Navg \sim 1$ and $\Navg \sim 2$.  As discussed by B03,
a halo that is only a few times $\Mmin$ usually ``spends'' its extra
baryons building a larger central galaxy, instead of making two
galaxies of comparable mass. A halo has to be $\sim$20 
times more massive than $\Mmin$ before it has a high chance of merging
with a halo massive enough to contribute a satellite above the threshold.

For N-body subhalos analyzed in K04, the scaling factor between $\Mmin$ and 
$M_1$ is about 25 at $z=0$ (and becomes smaller at higher $z$, e.g., 
$\sim$10 at $z=3$). The modest difference between our results and theirs 
may be caused partly by slight differences in defining $\Mmin$ and 
partly by the difference in the cosmological models adopted in the 
calculation. K04 also find that the high mass slope ($\alpha$)
of the satellite mean occupation is very close to one, independent of the 
mass threshold. We find similar results through our fits to the satellite 
occupation functions for galaxy samples with different baryonic mass 
thresholds (see Table~\ref{tbl:params}).

\section{Relations between Galaxy Mass/Luminosity and Halo Mass}
\label{sec:clf}

\subsection{Conditional Mass/Luminosity Function as a Function of Halo Mass}

The conditional luminosity function (CLF; \citealt{Yang03}) is the 
luminosity function $\Phi(L|M)$ of galaxies that reside in halos of a given 
mass $M$.  The CLF encodes the luminosity dependence of the mean
occupation function, since a complete characterization of $\Phi(L|M)$
allows one to examine the $M$-dependence for any specified range of $L$.
This formalism has been used to model observations of the 2dFGRS
and the DEEP2 redshift surveys by simultaneously fitting measurements
of the galaxy luminosity function and luminosity dependent clustering
(\citealt{Yang03,Bosch03a,Bosch03c,Yan03}). 
A derived or assumed CLF can be used to construct mock catalogs,
allowing more detailed tests of the CLF and cosmological model 
(e.g., \citealt{Mo04,Yan04,Yang04}).
The spirit of CLF modeling is similar to that of fitting parameterized
HOD models like those of \S\ref{sec:hod} to observed clustering in
different luminosity ranges, as undertaken for the SDSS by \cite{Zehavi05} 
(also see \citeauthor{Seljak05} [\citeyear{Seljak05}] for an analysis
using galaxy-galaxy lensing measurements).  In HOD fits, information about 
the galaxy luminosity function enters through the number density constraints.
The CLF approach attempts to solve for the entire luminosity dependence of 
the HOD at once, with the consequence that it requires a more detailed
parameterized form for model fitting.  The above mentioned papers 
usually assume that the CLF follows a \cite{Schechter76} function at each 
halo mass, and they parameterize the halo mass dependence of the 
normalization, faint end slope, and characteristic luminosity of
this Schechter function with various functional forms.  The central galaxy 
in each halo is chosen to be the most luminous one of the galaxies drawn from 
the CLF.
 
Since our galaxy sample is constructed based on a baryonic mass threshold,
here we first examine the theoretical predictions for conditional galaxy 
baryonic mass functions, which we refer to as the CMF (note that $M$ here 
represents galaxy baryonic mass, not halo mass).  The CMF should be similar to 
the CLF, though the systematic mass dependence and scatter of stellar
mass-to-light ratios will stretch and smooth the CLF somewhat relative
to the CMF.  The qualitative results for the SPH simulation 
are similar to those for the SA model except for the overall shift in baryonic 
mass scale. We concentrate on the SA model because it has better 
statistics and because it better matches the observed galaxy luminosity
function.  The SA model also predicts galaxy luminosities, and 
\citet{Benson03b} have presented SA calculations of CLFs. 
We compare CLF and CMF results later in this section.

Figure~\ref{fig:clf_fit} plots the CMF in bins of halo mass running
from $\log M=11.6 \pm 0.1$ to $\log M = 14.6 \pm 0.1$.  Solid
histograms show the full CMF, dashed histograms the contribution
of central galaxies only, and points with Poisson error bars the
contributions of satellite galaxies only.  Since we consider only
galaxies with baryonic mass $\Mgal > 1.45 \times 10^{10}\hMsun$, the
CMF in halos of mass $M \la 10^{12}\hMsun$ is completely dominated
by central galaxies; the two contributions become equal at halo mass
$M\approx 10^{13}\hMsun$, where $\Nsatavg=1$ (Fig.~\ref{fig:MminM1}).
The central galaxy CMF is sharply peaked at a fixed halo mass,
and it can be approximated by a Gaussian in $\log \Mgal$.  Solid lines
in Figure~\ref{fig:clf_fitpar} show the parameters of such Gaussian
fits as a function of $M$.  The dispersion is $\sigma\approx 0.13$
over a wide range, and much of this width reflects the $\pm 0.1$
size of our $\log M$ bins.  The dispersion rises at low and high
halo masses.  At low $M$, the mean central
galaxy mass is linearly proportional to halo mass, while the relation at 
high mass is much shallower, approximately $\Mgal \propto M^{1/3}$.
The ratio of the integrated central galaxy CMF to the total CMF
follows $\Ncenavg/(\Ncenavg+\Nsatavg)$ once the baryonic mass scale of the 
central galaxies is well above the threshold baryonic mass, which happens
in halos more massive than $\sim 2\times 10^{12}\hMsun$.

The satellite galaxy CMF rises monotonically towards lower $\Mgal$
at any halo mass, and it can be approximated by a Schechter function
that is truncated at the middle of the Gaussian representing the
central galaxy CMF.  Thin solid curves in Figure~\ref{fig:clf_fit}
show a model in which the satellite contribution is a sharply
truncated Schechter function with $\alpha_s=-1.5$ and 
$\Mgal^*=10^{11}\hMsun$ at every halo mass\footnote{Note that we use
$\alpha_s$ to distinguish the Schechter function slope from the slope
$\alpha$ of $\Nsatavg$.} and the central galaxy
contribution is a Gaussian in $\log \Mgal$ with constant width 
$\sigma = 0.13$. The mean of the Gaussian, shown by the dashed line in the 
middle panel of Figure~\ref{fig:clf_fitpar}, is 
\begin{equation}
\label{eqn:meanmgal}
\mu(M) = \mu_t \left[\frac{1}{1+c}(M/M_t)^{-\gamma}+\frac{c}{1+c}(M/M_t)^{-\gamma/3}\right]^{-1/\gamma},
\end{equation}
where $\mu_t=5.9\times 10^{10}\hMsun$, $M_t=2.3\times 10^{12}\hMsun$, $c=0.49$,
and $\gamma=5.8$.
The normalization of the truncated Schechter function is determined by 
matching the average number of satellite galaxies above the baryonic 
mass threshold in each halo mass bin, and the central galaxy
CMF integrates to unity by definition (i.e., we assume that all halos
contain a central galaxy of some mass, though it might be below our
adopted threshold). The top panel of Figure~\ref{fig:clf_fitpar} shows the 
relative amplitudes of central and satellite CMFs evaluated at the mean of 
the Gaussian component. Even though the relative normalization is the
only parameter adjusted on a bin-by-bin basis, this global fit describes
the SA model CMF quite well over the full range of halo masses
plotted in Figure~\ref{fig:clf_fit}. 

With the CMF/CLF, one can populate halos from simulations to study various
properties of galaxy clustering. For a halo of a given mass, one chooses the
baryonic mass or luminosity for the central galaxy according to the central
galaxy CMF/CLF and does similarly for the satellites. This procedure 
implicitly assumes that there is no correlation between the
masses/luminosities of central galaxies and satellites in halos of 
fixed mass.  Figure~\ref{fig:ccmf} tests the validity of this assumption by
comparing satellite CMFs at different values of central galaxy baryonic
mass in three narrow bins of halo mass.  The central and satellite CMFs
are normalized so that there is one central galaxy per halo.
For $10^{13} M_\odot$ halos, there is a clear trend for a lower 
amplitude satellite CMF in halos with a more massive central galaxy.
The trend is present but weak in the more massive halos.
This trend could be a consequence of galactic cannibalism, with
more massive central galaxies growing by consuming more satellites.
Note that in a small fraction of halos, the most massive satellite
exceeds the mass of the central galaxy.  While the SA model always
places the more massive galaxy from a pair of merged halos at the
center of the new halo, that galaxy may be more gas rich than its
partner, in which case feedback from star formation can leave it
slightly lower mass in the end.

A clear consequence of the central galaxy contribution is that the CMF
at a given halo mass {\it cannot} be well approximated by a Schechter
function, especially at intermediate masses 
$10^{12.5}\hMsun \la M \la 10^{14}\hMsun$ where the central galaxy
``bump'' rises well above the extrapolation of the satellite CMF.
At high $M$ the fractional contribution of the central galaxy
is small, and a Schechter function is a reasonable approximation
for most purposes.  At low $M$, where central galaxies dominate, the Gaussian 
CMF can be roughly approximated by a Schechter function with
$\alpha_s > 0$.  \cite{Yang03} indeed find $\alpha_s > 0$ at low masses
in their (Schechter-based) fits to the 2dFGRS conditional luminosity
function (see their eq.~[20] and Table~1).  Note, however, that if
we lowered our galaxy mass threshold we would pick up lower mass
(fainter) satellites in low mass halos, and a Schechter form would again 
become a poor description. 

\citet{Benson03b} study the CLF in SA models and find a 
similar trend to that seen here in the CMF: in low mass halos the CLF has a 
``hump'' at the bright end and a roughly power-law shape at the faint end, 
and in massive halos the ``hump'' disappears and the CLF approaches the 
Schechter form  (see their Fig.~1). 
Figure~\ref{fig:clf_gr} presents similar results for the cosmology
and version of GALFORM adopted here, in Sloan $g$ and $r$ bands.
We truncate the histograms at $M_g=-18.75$ and $M_r=-19.25$,
since scatter in the mass-luminosity relation would make our
mass-limited galaxy samples incomplete at fainter magnitudes.
The central galaxy hump is much broader than in the CMF, and it
therefore merges more continuously into the total CLF, though it
is still visible in the low mass histograms.
While the total CLF roughly resembles a Schechter function with
a faint-end slope that changes steadily with halo mass, our
results suggest that a better model for empirical fitting
would be a suitably generalized version of the
one used in Figure~\ref{fig:clf_fit}, with parameterized forms
of $\Nsatavg/\Ncenavg$ and $\mu(M)$, and freedom (perhaps
mass dependent) added to the width $\sigma$ of the central galaxy CLF
and the Schechter parameters of the satellite CLF.
\cite{Yang03} considered a model of this form in their first paper
on the CLF and showed that it yields a good fit to the 2dFGRS data,
though they have mostly adopted the Schechter form in their later work.

How does separating central and satellite galaxies change our view of
the overall galaxy mass function, and hence the luminosity function?
Figure~\ref{fig:lf_cen_sat}a shows the galaxy baryonic mass function (MF)
of the SA model and the individual contributions of central and
satellite galaxies.  Both of these contributions, and the total MF,
can be reasonably well fit by Schechter functions, even though the CMF
at fixed halo mass cannot.  Central galaxies dominate the total MF
at every mass; the satellite contribution is $\sim 37\%$ in our lowest
mass bin (corresponding to luminosity $\sim 0.18 L_*$) and $\la 18\%$
for $\Mgal \ga 10^{11}\hMsun$ ($L \ga 1.20L_*$).
The dominance of central galaxies is a consequence of the large
gap between the minimum halo mass $\Mmin$ for hosting a central
galaxy and the mass $M_1 \approx 18\Mmin$ required to host a satellite
of the same mass.  The more massive halos are rarer,
especially in the exponential cutoff region of the halo mass function,
so even though a massive halo can host multiple satellites, the
central galaxies of abundant, lower mass halos dominate by number.
Figure~\ref{fig:lf_cen_sat}b amplifies this point by dividing the MF
into contributions from different halo mass ranges.
In observational terms, most galaxies are found in group environments,
but these groups (at least if defined at an overdensity $\sim 200$)
typically have a central galaxy that is substantially brighter than
its neighbors.  If one chooses a random galaxy of a given luminosity
(especially with $L>L_*$), then it is more likely to be the dominant
galaxy of its own group rather than the satellite of a more luminous system.

Our results on the decomposition of the MF are in agreement with those found 
for the LF by \citet{Benson03b}; see their Fig.~3. 
They also find that, except for very low mass halos (where the
photoionization suppression of galaxy formation becomes important), the faint 
end slope of the CLF is steeper than that of the overall LF, 
$\alpha_s \sim -1.5$ versus $\alpha_s \sim -1.2$. The shallower faint end 
slope of the overall LF reflects the contribution of central galaxies. 
Our results for the galaxy MF are also similar to results found by other 
authors for the MF of dark matter subhalos in high resolution $N$-body 
simulations. \citet{Vale04}, based on the simulations of \citet{Weller04},
propose that the satellite subhalo mass function in a given halo can be fit 
by a Schechter function with low mass slope $\alpha_s\sim-1.91$,
which is consistent with the power-law slope $\sim -2$ found by 
\citet{Delucia04}. Compared with their results, we find a shallower low mass
slope ($\alpha_s\sim -1.5$), probably because of the difference between dark 
matter and baryonic mass.  Figure~1 of \citet{Vale04} shows the overall mass 
functions of (satellite) subhalos and parent halos, which conveys similar 
information to the lower left panel of our Figure~\ref{fig:lf_cen_sat}.

\subsection{Mass Distribution of Host Halos for Galaxies at Fixed 
Mass/Luminosity}

Orthogonal to the CMF/CLF is the conditional mass distribution of 
{\it halos} hosting galaxies at fixed baryonic mass or luminosity. 
This conditional distribution is just the product of the halo mass 
function and the mean occupation function of such galaxies (see 
Figure~\ref{fig:Mgalbin} for the mean occupation function for a sample 
of galaxies in a bin of baryonic mass).  This quantity is relevant to 
the interpretation of galaxy-galaxy lensing measurements or to other 
methods of estimating the average mass distribution surrounding galaxies 
of specified properties.
 
Figure~\ref{fig:PML} shows the SA model predictions for the conditional 
mass distribution of host halos in narrow bins of galaxy baryonic mass 
(top panels) and $r$-band luminosity (bottom panels). The distribution is 
separated into contributions from central and satellite galaxies as well 
as blue and red galaxies.  The division of blue and red galaxies is based 
on a color cut at $g-r=0.734$, which results in roughly equal numbers of 
blue and red galaxies.  For galaxies of fixed baryonic mass, the host halos 
span a wide range of masses, with a peak at low masses contributed by 
central galaxies and a fairly flat tail to high masses from satellites.
The host halos of red galaxies, whether central or satellite, tend to be 
more massive than the host halos of blue galaxies.  As the galaxy baryonic 
mass increases, the contributions to the conditional halo mass 
distribution from satellite galaxies and from blue galaxies decrease, 
as expected.  

As a function of galaxy luminosity, the conditional mass distribution of 
host halos has a trend similar to that with baryonic mass.  However,
the scatter between galaxy luminosity and baryonic mass broadens the
distribution, especially the central galaxy peak, and it leads
to greater separation between the typical halo masses of blue and red 
central galaxies.  This result is analogous to the finding of
\citet{Mandelbaum05}, who assign luminosities to subhalos identified in 
high-resolution dissipationless simulations and show that scatter between 
subhalo circular velocity and galaxy luminosity makes the halo mass 
distribution at fixed luminosity wider.  The width and asymmetry of the 
conditional mass distributions implies, first, that the mean halo mass 
estimated for a given type of galaxy can be very different from the 
typical mass of an {\it isolated} halo that hosts such a galaxy (i.e., 
from the location of the central galaxy peak in these histograms).  Second, 
it suggests that interpretation of galaxy-galaxy lensing measurements will 
be simpler if one can classify galaxies by baryonic mass instead of 
luminosity and if one can separate isolated galaxies from those that are 
likely to be satellites in more massive halos.

\section{Summary and Discussion}
\label{sec:disc}

A number of previous studies have investigated the halo occupation function
$P(N|M)$ predicted by semi-analytic models, hydrodynamic simulations,
and high-resolution N-body simulations
(\citealt{Kauffmann97,Benson00,Seljak00,Scoccimarro01,Sheth01,White01,
Yoshikawa01,Cooray02,Guzik02,Scranton03}; B03; K04),
showing good agreement in the qualitative
features predicted by the different methods.  Here we have followed
the lead of \cite{Guzik02} and K04 by separating the
contributions of central and satellite galaxies to $P(N|M)$.
When we consider galaxy samples defined by thresholds in baryonic mass,
analogous to observational samples defined by thresholds in luminosity,
our results for the \cite{Cole00} SA model and the \cite{Dave02}
SPH simulation are similar to those found by K04 for subhalos in
high resolution N-body simulations.  In particular, the separation of
central and satellite galaxies naturally explains the general shape
of the mean occupation function $\Navg$ and the transition from
sub-Poisson fluctuations in $P(N|M)$ at low $N$ to roughly Poisson
fluctuations at high $N$.  The correlation between halo mass and
central galaxy mass is tight, so $\Navg$ for central galaxies rises
sharply at a threshold mass $\Mmin$ and can be approximated by a
step function.  A halo has either zero or one central galaxies, so
the width of $P(N_{\rm cen}|M)$ is sub-Poisson by definition.
The mean occupation of satellites has a roughly power law form,
$\Nsatavg \approx (M/M_1)^\alpha$ with $\alpha \approx 1$,
and the fluctuations of $P(N_{\rm sat}|M)$ about the mean are
close to Poisson.  However, a halo must be $\sim 10-20\Mmin$ before
it hosts on average one satellite galaxy above the baryonic threshold; in the
mass decade above $\Mmin$, a larger halo typically hosts a larger
central galaxy instead of multiple galaxies above the threshold (B03).
In the halo mass regime where central galaxies make a significant
contribution to the total galaxy counts, the mean occupation rises
slowly, and the width of the total $P(N|M)$ is substantially narrower
than a Poisson distribution.

When a sample of galaxies above a baryonic mass threshold
is divided in two on the basis
of stellar population age, the HODs of young and old galaxies are
markedly different.  Halos near the cutoff mass $\Mmin$ tend to host
young central galaxies, while more massive halos host old central
galaxies.  The mean occupation function of young galaxies exhibits
a local minimum at halo masses $M \sim 10\Mmin$, where halos are
too massive to have a young central galaxy but not massive enough
to have satellites.  The old galaxy population has a monotonically
rising $\Navg$, and the number of old satellites in massive
halos is larger than the number of young satellites.  The statistics of
old and young satellites are consistent with Poisson distributions
about their respective means, uncorrelated with the age of the central
galaxy.  If one starts with a sample of galaxies in a bin of baryonic
mass (instead of a mass-threshold sample), then the halos that
host central galaxies occupy a relatively narrow mass range, and the ratio
of old central galaxies to young central galaxies depends on the
galaxy mass bin.  For a low mass (or low luminosity) sample, most central 
galaxies are young, and the sample's
older galaxies are therefore satellites in higher mass halos.
With a higher mass (or luminosity) sample, more of the old galaxies
are central objects.  These differences in the HODs of young and old
galaxies naturally explain much of the observed dependence of galaxy
clustering on color, spectral type, and morphology, with red/early-type
galaxies exhibiting stronger correlations and residing in higher
density environments.

Two striking features of the theoretically predicted HODs are the
near-constant ratio and large gap between the minimum host halo mass
$\Mmin$ and the mass $M_1$ at which an average halo hosts one satellite
above the baryonic mass threshold.
The ratio is $M_1/\Mmin \approx 14$ for the SPH simulation and
$M_1/\Mmin\approx 18$ for the SA model, over a wide range in galaxy mass.
The large value of $M_1/\Mmin$ accounts for the extended plateau
in $\Navg$ at low halo masses, and it has a number of important
consequences.  First, the conditional (galaxy) mass function (CMF) at
fixed halo mass is not well described by a Schechter function.  Central 
galaxies produce a ``bump'' in the CMF at high galaxy masses that rises well 
above the Schechter extrapolation of the satellite population.  
Scatter in the relation between baryonic mass and luminosity smears
out but does not eliminate this bump in the conditional luminosity function
(CLF).  Second, at any given galaxy mass or luminosity, the total 
mass/luminosity function is dominated by central galaxies, because the 
massive halos that can host multiple satellites are much less abundant than 
halos with $M \sim \Mmin$.  The satellite fraction is larger at low
masses/luminosities, where halos with $M \sim M_1$ are not yet on
the exponential tail of the halo mass function.

The large value of $M_1/\Mmin$ also plays a key role in shaping the
galaxy correlation function $\xi(r)$.  As first emphasized by
\cite{Benson00}, and confirmed in many subsequent investigations
(e.g., \citealt{Peacock00,Seljak00,Scoccimarro01,Berlind02}), the
sub-Poisson width of $P(N|M)$ at low masses is crucial to
reproducing the observed, roughly power law form of $\xi(r)$, because
it enables low mass halos to host large numbers of galaxies without
hosting large numbers of small separation pairs.  As shown here and
in K04, this sub-Poisson width is a consequence of the nearest-integer
statistics of central galaxy occupations, and it holds over an extended
range of halo masses because $M_1/\Mmin$ is large.  The roughly
constant value of $M_1/\Mmin$ enables galaxies with a wide
range of luminosity to exhibit power law correlation functions.
In detail, HOD models can also explain the deviations of $\xi(r)$
from a power law found in high precision measurements, as shown by
\cite{Zehavi04}.

The nearly constant value of $M_1/\Mmin$ and the nearly Poisson form
of $P(\Nsat|M)$ at fixed $M$ are pleasingly simple results, and in
qualitative terms they seem intuitively sensible.
However, we do not have a quantitative explanation for either one of them.
An important clue is that K04 find essentially the same results
for samples of dark matter subhalos defined by circular velocity
thresholds in purely gravitational simulations, which suggests that
they emerge mainly from dark matter dynamics and that gas physics
and star formation serve to light up the dark matter substructures in
a fairly straightforward fashion.  The $M_1/\Mmin$ scaling and form
of $P(\Nsat|M)$ should then be determined mainly by the statistics of halo
merger histories, though they will also be affected by tidal disruption
of subhalos and galaxies and by mergers induced by dynamical friction.
Analytic methods have been
developed to model these processes \citep{Bond91,Bower91,Lacey93,Bullock00,
Taylor04}. A recent theoretical study of the formation and evolution of 
substructures using the methods of \citet{Zentner05}
shows that tidal stripping and disruption are the most 
important mechanisms shaping the HOD, with dynamical friction sub-dominant
(A. Kravtsov, private communication). Further investigation along these lines
may lead to a more fundamental theory of the galaxy HOD and generic
predictions for its dependence on redshift and on cosmological parameters.

Recent observational analyses show qualitative and to some degree
quantitative agreement with many of our theoretical predictions.
\cite{Zehavi05} show that the projected correlation functions of
SDSS galaxy samples with a wide range of luminosity thresholds can be
well fit by the 3-parameter ($\Mmin$, $M_1$, $\alpha$) HOD models
described in \S\ref{sec:hod_all} for ``concordance'' values of
cosmological parameters.  The fitted values of $\Mmin$ and $M_1$
show the predicted linear scaling relation, with $M_1 \approx 23 \Mmin$,
and the dependence of $\Mmin$ on luminosity resembles the
predicted dependence on galaxy baryonic mass
(compare Zehavi et al.'s Fig.~18 to our Fig.~\ref{fig:MminM1}).
The measured ratio $M_1/\Mmin \approx 23$ is higher than the value of 14
predicted by the SPH simulation, probably for the same reasons that
the SPH simulation fails to match the observed galaxy luminosity function.
The SA model is designed to match the luminosity function fairly well,
and its predicted ratio of 18 is closer to the measured value (the factor
$\sim$25 in the dissipationless simulations of K04 is closer still).
The remaining discrepancy could largely reflect the difference between 
a luminosity-threshold and mass-threshold galaxy sample, since scatter in 
stellar mass-to-light ratios allows some galaxies of the 
luminosity-threshold sample to occupy lower mass halos, reducing $\Mmin$
and simultaneously raising $M_1$ to keep the number density fixed. 
The difference in cosmological models 
(we assume higher $\Omega_m$ and lower $\sigma_8$ than Zehavi et al.) and 
errors in deriving $\Mmin$ and $M_1$ from the clustering data with a 
simplified HOD model can also contribute to the discrepancy.

\begin{figure}[h]
\plotone{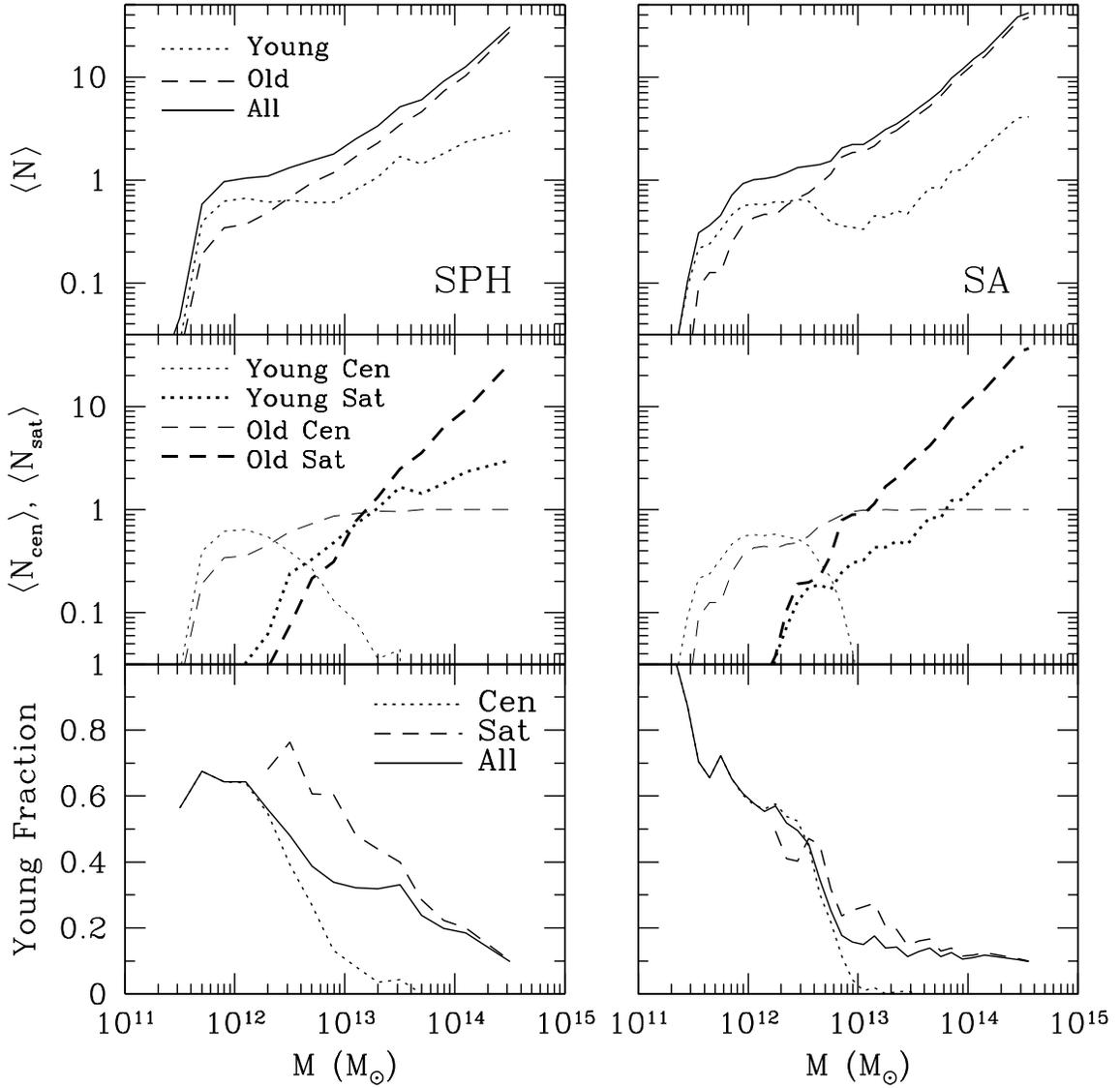}
\caption[]{\label{fig:color_hod}
Age dependence of the HOD predicted by the SPH simulation (left panels)
and by the SA model (right panels). For each model,
the mean occupation functions of old and young galaxies are shown in the top 
panel, contributions from central and satellite galaxies to the mean 
occupation number are plotted in the middle panel, and the fraction of young 
galaxies (in central, satellite, and all galaxies) is plotted in the bottom 
panel. 
}
\end{figure}

\begin{figure}
\plottwo{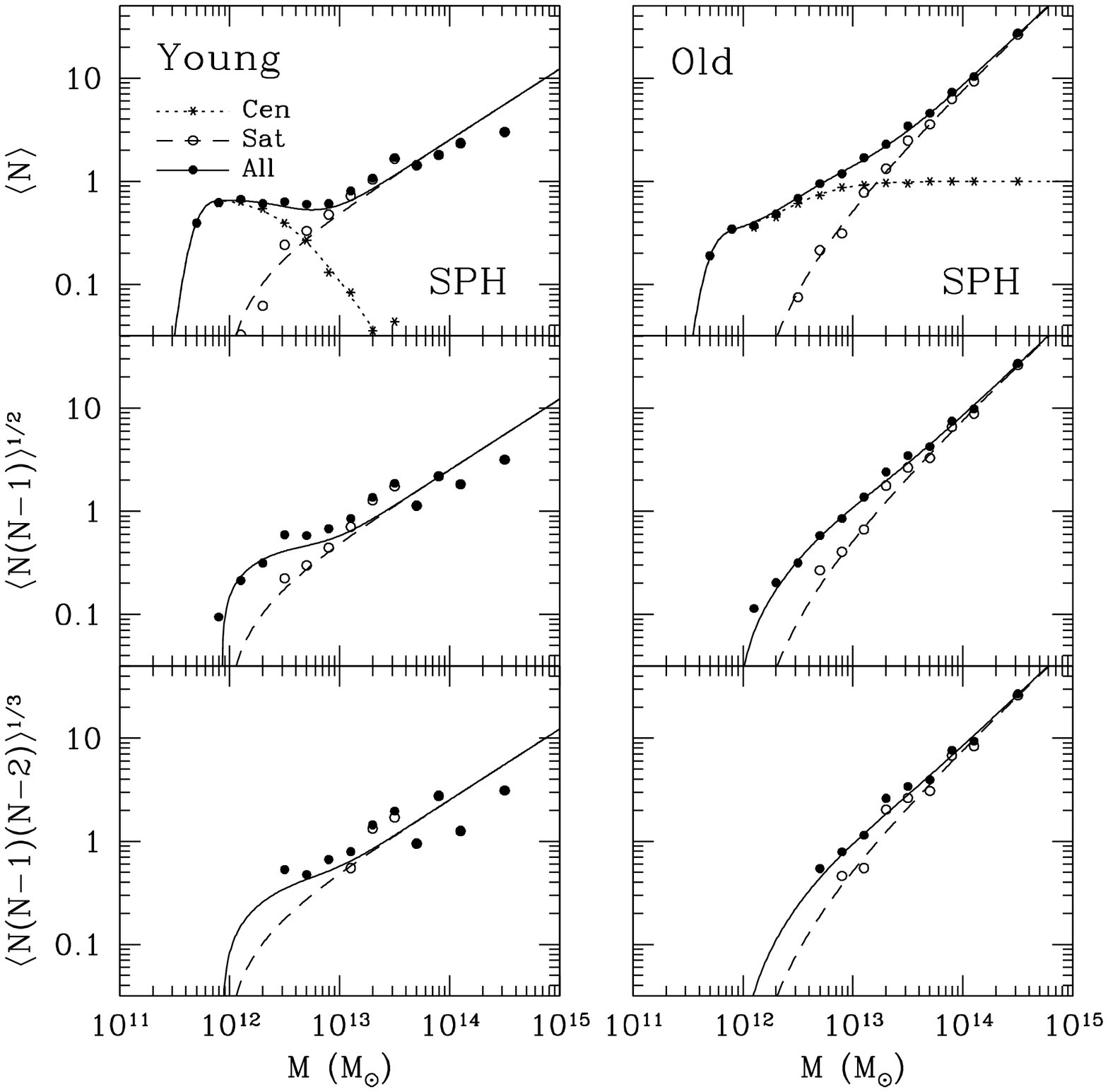}{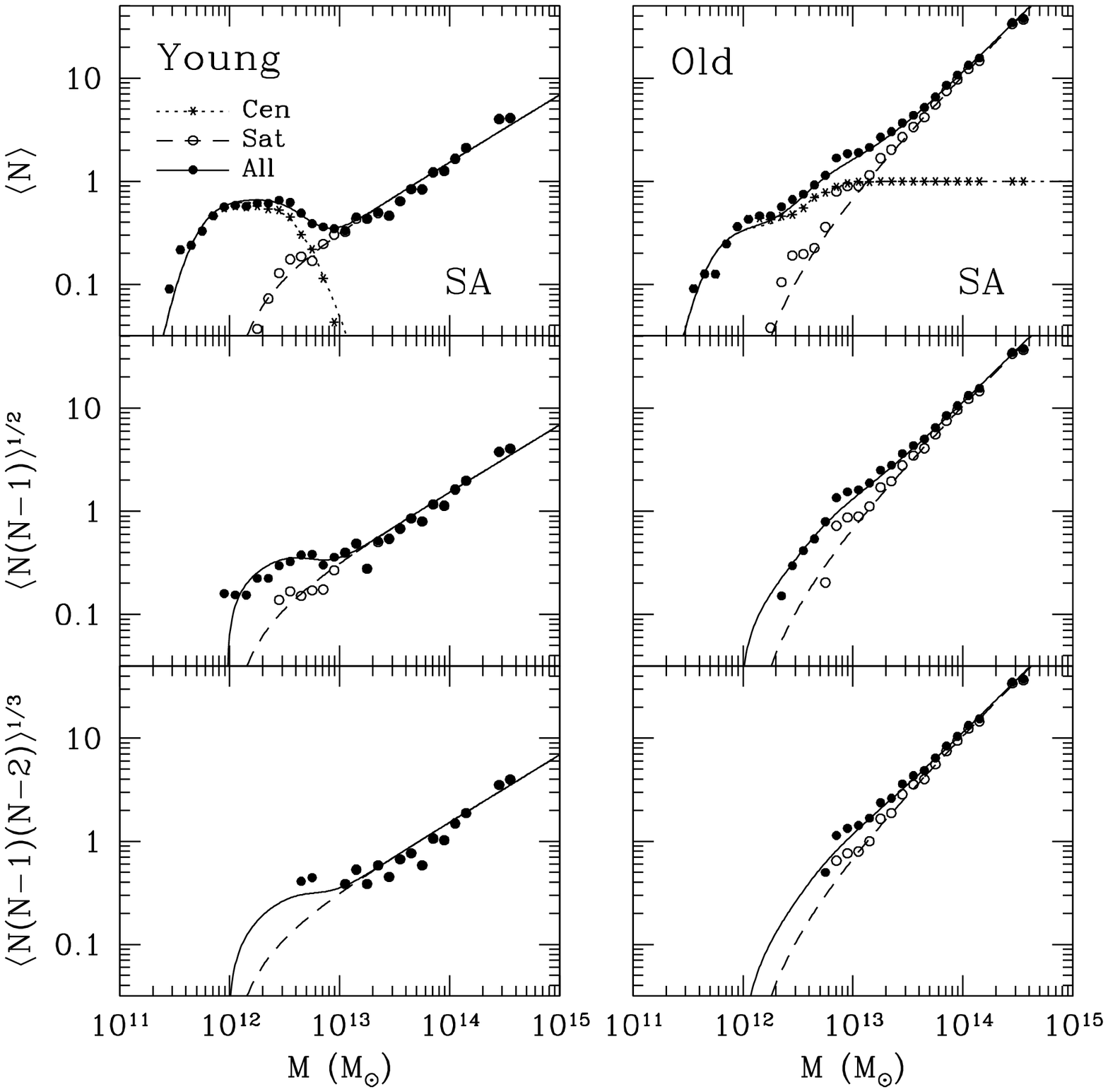}
\caption[]{\label{fig:parfit_c}
Same as Fig.~\ref{fig:parfit}, but for each model the left panels show
the young galaxy sample and right panels the old galaxy sample. Points show
the model results and lines show fits using the 5-parameter model for all
galaxies and the blue fraction parameterization described in the text.
}
\end{figure}

\cite{Zehavi05} also find that the clustering of red and blue
subsamples can be well described by HOD models like those described
in \S\ref{sec:hod_age} for old and young galaxy populations.
Fits to the blue fraction of central and satellite galaxies
as a function of halo mass show good qualitative agreement with our
predictions (compare Zehavi et al.'s Fig.~22 to our Fig.~\ref{fig:color_hod}).
Van den Bosch, Yang, \& Mo (\citeyear{Bosch03b}) carry out CLF fits to
the 2dFGRS data, and their inferred dependence of late-type galaxy fraction 
on halo mass is qualitatively similar to our prediction for blue galaxy 
fraction.  Except for the lowest luminosity bin, their inferred mean 
occupation function for luminosity-bin galaxy samples (see their Fig.~10)
does not show a clear bump like that seen in our Figure~\ref{fig:Mgalbin},
probably because of their assumed Schechter function form of the CLF 
(see below). 

\cite{Zehavi05} use their HOD fits as a function of luminosity to
infer the conditional luminosity function in different ranges of
halo mass.  Given the good agreement with the predicted HOD results,
it is not surprising that their inferred CLFs resemble our predicted
conditional mass functions (compare their Fig.~21 to our
Fig.~\ref{fig:clf_fit}).  In particular, the large $M_1/\Mmin$
ratio leads to a central galaxy ``bump'' in the luminosity function
of intermediate mass halos.  Indeed, these bumps are more prominent than 
those seen in our conditional {\it luminosity} function predictions 
(Fig.~\ref{fig:clf_gr}), perhaps indicating that the SA model used here 
produces too much scatter in the baryonic mass-luminosity relation for 
central galaxies.  \citet{Hansen04} have measured
the CLF directly in an SDSS cluster catalog \citep{Bahcall03} and
find that bumps caused by the brightest cluster galaxies emerge in the 
CLFs of low richness clusters. They also find that, as a result of the 
significant contributions from the brightest cluster galaxies, a Schechter 
function is not a good fit for the CLFs in many of the cluster richness 
bins.  \citet{Eke04} look for such central galaxy bumps in the CLF of groups 
and clusters in the 2dFGRS Percolation-Inferred Galaxy Group (2PIGG) catalog 
and do not find them, but they demonstrate using SA mock catalogs that 
measurement errors in the group masses would wash them out. Using a 
halo-based group finder (\citealt{Yang05a}), \citet{Yang05b} identify groups 
from the 2dFGRS and measure the HOD as a function of halo mass. They find a 
tight correlation between the mean luminosity of central galaxies and the 
halo mass, qualitatively similar to that shown in the central panel of 
Figure~\ref{fig:clf_fitpar} of this paper. Although they obtain reasonably 
good fits to the CLF over the halo mass range 
$13\lesssim \log[M/(h^{-1}M_\odot)] \lesssim 14.5$
with a Schechter function, the bright end of the CLF in halos 
of $\log[M/(h^{-1}M_\odot)]\sim 13$ starts to show a clear enhancement caused 
by the central galaxies. \citet{Lin04a} and \citet{Lin04b}
find that the $K$-band LF of 2MASS cluster galaxies can be fit by a         
Schechter function if the brightest cluster galaxies are excluded. 
The brightest cluster galaxies seem to follow a different distribution and 
become less important in the total cluster light as the cluster mass 
increases. They also find that the distribution of satellite numbers at a fixed 
estimated mass is roughly Poisson. \citet{Miles04} study a sample of 25
groups drawn from the Group Evolution Multi-wavelength Study (GEMS), and they
find prominent bumps at the bright end of the $B$-band and $R$-band LFs of
the group galaxies, which are very likely caused by central galaxies. 
The isolated, luminous, X-ray bright ellipticals known as ``fossil groups'' 
(\citealt{Vikhlinin99,Jones03}) may represent an extreme example of the 
central galaxy bump in intermediate mass halos.
All of these results are in good qualitative agreement with the
predictions here, though more careful replication of the observational
selection would be needed for quantitative comparison.

In our decomposed CMFs and CLFs (Figs.~\ref{fig:clf_fit}
and~\ref{fig:clf_gr}), the satellite contributions are fairly
well described by a truncated Schechter function with constant
faint-end slope $\alpha_s$, but the shape of the total CMF/CLF
changes with halo mass because of the changing relative amplitude of
central and satellite contributions.  \cite{Yang03} infer the CLF from 
2dFGRS data {\it assuming} a Schechter form at each halo mass, and they 
find a steadily changing $\alpha_s$ that becomes positive 
in low mass halos, which might plausibly be a consequence of describing
the roughly Gaussian luminosity function of the central galaxies with
a Schechter function.  The Schechter+Gaussian
model described in \S\ref{sec:clf} should be a better parameterized
form for fitting observational data. It is not clear how forcing
a Schechter function fit might influence the {\it cosmological} conclusions
from the CLF method \citep{Yang03,Bosch03c}, but we would expect some impact.
For example, a Schechter-based CLF fit that gives the correct halo
occupations of $L_*$ galaxies will give incorrect occupations, and
hence incorrect bias factors, for $2L_*$ galaxies, and thus alter the 
inferred amplitude of mass clustering. \cite{Yang03} investigate an 
alternative parameterization for the CLF similar to that advocated here:
they assume the central galaxy CLF to be a log-normal function for 
low-mass halos, with the width being a free parameter. With this alternative 
parameterization and a concordance cosmological model, they find that
the inferred width in the central CLF is more or less consistent with the 
scatter in the Tully-Fisher relation, and their fits to the global luminosity 
function and the luminosity dependence of the correlation length do not change 
much (see their Fig.~7). The resulting best-fit HOD has noticeable changes, 
especially at the low halo mass end.  This test suggests that the assumed CLF 
form can have a significant impact on inferred halo galaxy populations
but may have little effect on cosmological constraints derived from 
the galaxy luminosity function and luminosity dependence of 
galaxy clustering strength.

The qualitative agreement between our predictions and the results of
\cite{Zehavi05} suggests that current theoretical models of galaxy
formation are on largely the right track.  There should be interesting
progress in the near future as more clustering measurements are
incorporated into the observational determinations of the HOD.
For example, the group multiplicity function places fairly direct
constraints on $P(N|M)$ at high halo masses
\citep{Peacock00,Marinoni02,Berlind02,Kochanek03},
and void statistics and galaxy scaling relations are sensitive
to behavior in the cutoff region near $\Mmin$ \citep{Berlind02},
so combinations of these measures with the correlation function can constrain
more flexible parameterizations of the HOD.
Our 5-parameter model provides a near-perfect description of the
theoretical results and offers a good starting point for fitting observations,
but ultimately one would like to test the ingredients of this model
by allowing more general forms for $\Navg$, non-Poisson statistics
for satellites, and so forth.  Direct measurements of galaxy profiles
in groups and clusters can refine the standard assumption
(supported by the simulation results of B03) that satellite galaxies
trace the underlying dark matter distribution within halos.
Detailed empirical determinations of the HOD for different classes
of galaxies will test theoretical models of galaxy formation in
greater detail than previously possible, and they may provide
guidance to additional physics that should be incorporated in these models.

A still more ambitious goal is to constrain cosmological parameters
and the HOD simultaneously using observed galaxy clustering, either
on its own or in combination with other observables.
Based on the CLF approach, \citet{Bosch03c} have demonstrated
that the galaxy luminosity function, luminosity dependence of the galaxy
correlation length, and mass-to-light ratios of galaxy clusters can 
impose interesting constraints on cosmological parameters, 
either on their own or in combination with CMB data.
\citet{Abazajian05} combine CMB data with the projected correlation 
function of SDSS galaxies brighter than $\Mr=-21$, adopting parameterized 
HOD models like those in \S\ref{sec:hod_all}.  Cosmological constraints 
from this combination of measurements are as tight as those found by
combining CMB data with the large scale galaxy power spectrum
(e.g., \citealt{Percival03,Spergel03,Tegmark04}), where instead of
an HOD model one assumes that galaxy bias is scale-independent in
the linear and near-linear regime.  \citet{Tinker05} show that the SDSS 
correlation function constraints and observational estimates of cluster 
mass-to-light ratios imply that the values of $\Omega_m$ and/or $\sigma_8$ 
are lower than the traditionally adopted ``concordance'' values of 0.3 and 
0.9, in agreement with \citeauthor{Bosch03c}'s (\citeyear{Bosch03c}) 
conclusion based on CLF analysis of the 2dFGRS.  The HOD approach will become
more powerful as more clustering measurements are incorporated,
especially observables like redshift-space distortions and galaxy-galaxy
lensing that are directly sensitive to mass.  Even with a very flexible
HOD parameterization, galaxy clustering measurements alone can pin down
cosmological parameters like $\Omega_m$ and $\sigma_8$ with reasonable
precision (Z.~Zheng and D.~Weinberg, in preparation).  In concert with
CMB anisotropy, the Ly$\alpha$ forest, Type Ia supernovae, and other
observables that probe different scales and redshifts, careful modeling
of low redshift galaxy clustering allows sharpened tests of the nature
of dark energy, the masses of neutrinos, and the spectrum of density
and gravity wave fluctuations that emerged from the early universe.

\acknowledgments
We thank Andrey Kravtsov and Idit Zehavi for helpful conversations. 
We thank Xiaohu Yang, Frank van den Bosch, and Houjun Mo for useful 
comments on an earlier draft of the paper. We also thank the referee, 
Adi Nusser, for constructive comments. We thank Chris Kochanek for
suggesting the analysis that led to Figure~\ref{fig:PML}.
Z. Z. was partly supported by a Presidential Fellowship from the Graduate 
School of the Ohio State University. Z. Z. also acknowledges the support of 
NASA through Hubble Fellowship grant HF-01181.01-A awarded by the Space 
Telescope Science Institute, which is operated by the Association of 
Universities for Research in Astronomy, Inc., for NASA, under contract 
NAS 5-26555. This research was also supported by NSF grants AST 00-98584, 
AST 02-05969, and AST 04-07125, and NASA ATP grant NAGS-13308. C. G. L. 
acknowledges support from the PPARC rolling grant for extragalactic astronomy 
and cosmology at Durham. A. J. B. and C. M. B. are supported by Royal Society 
University Research Fellowships.

\input table.tex

\clearpage

\begin{figure}
\plotone{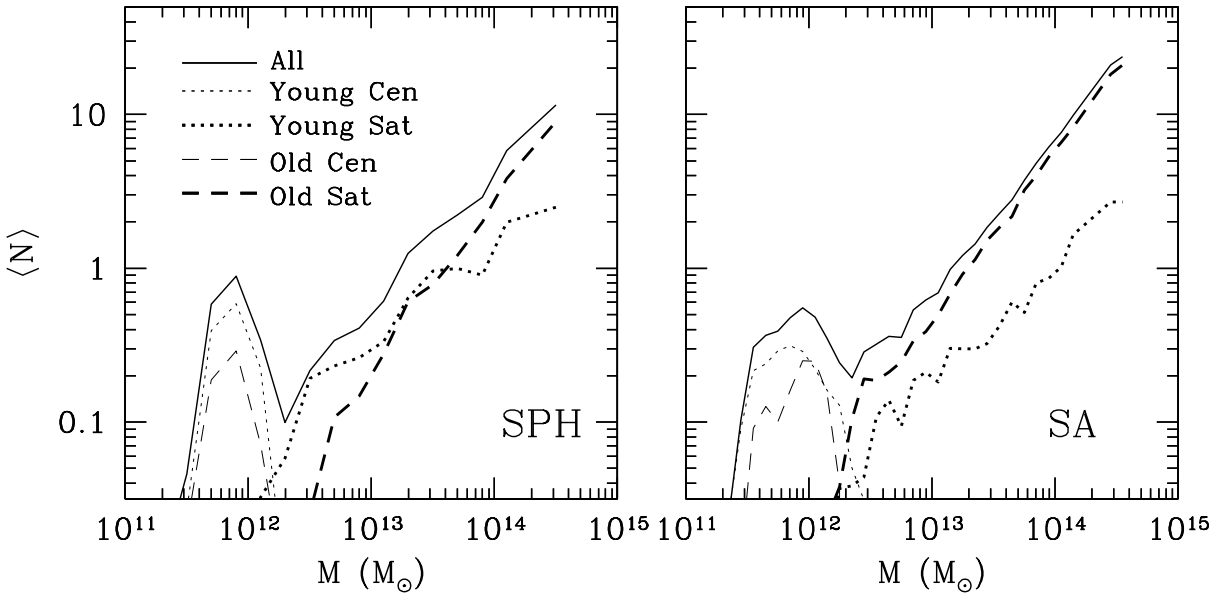}
\caption[]{\label{fig:Mgalbin}
HOD predicted by the SPH simulation (left panel) and by the SA model
(right panel) for a sample of galaxies in a bin of baryonic mass, containing
the less massive half of the full galaxy sample. Dotted and dashed curves 
show young and old galaxy contributions, respectively. 
}
\end{figure}

\begin{figure}[h]
\plottwo{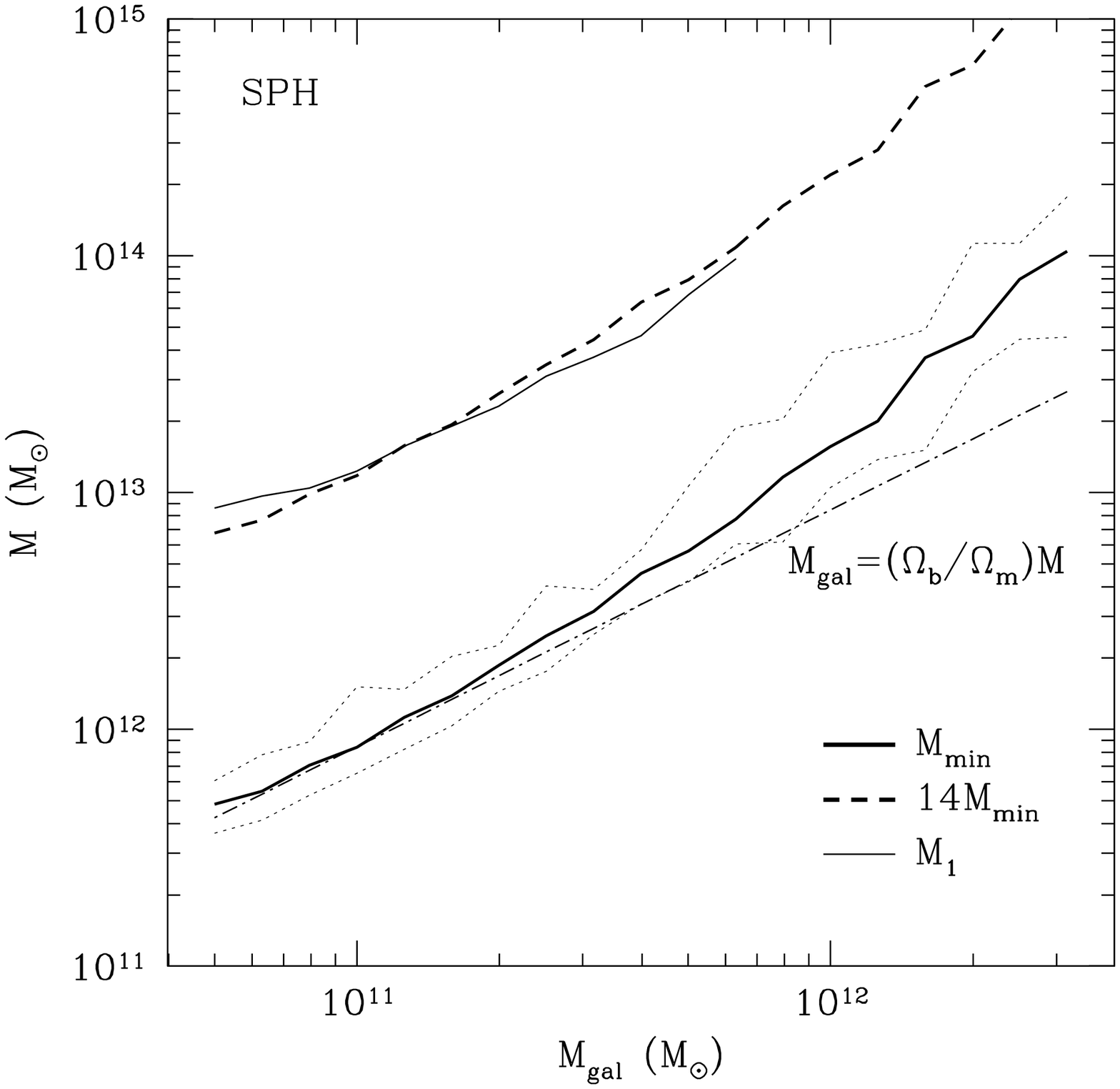}{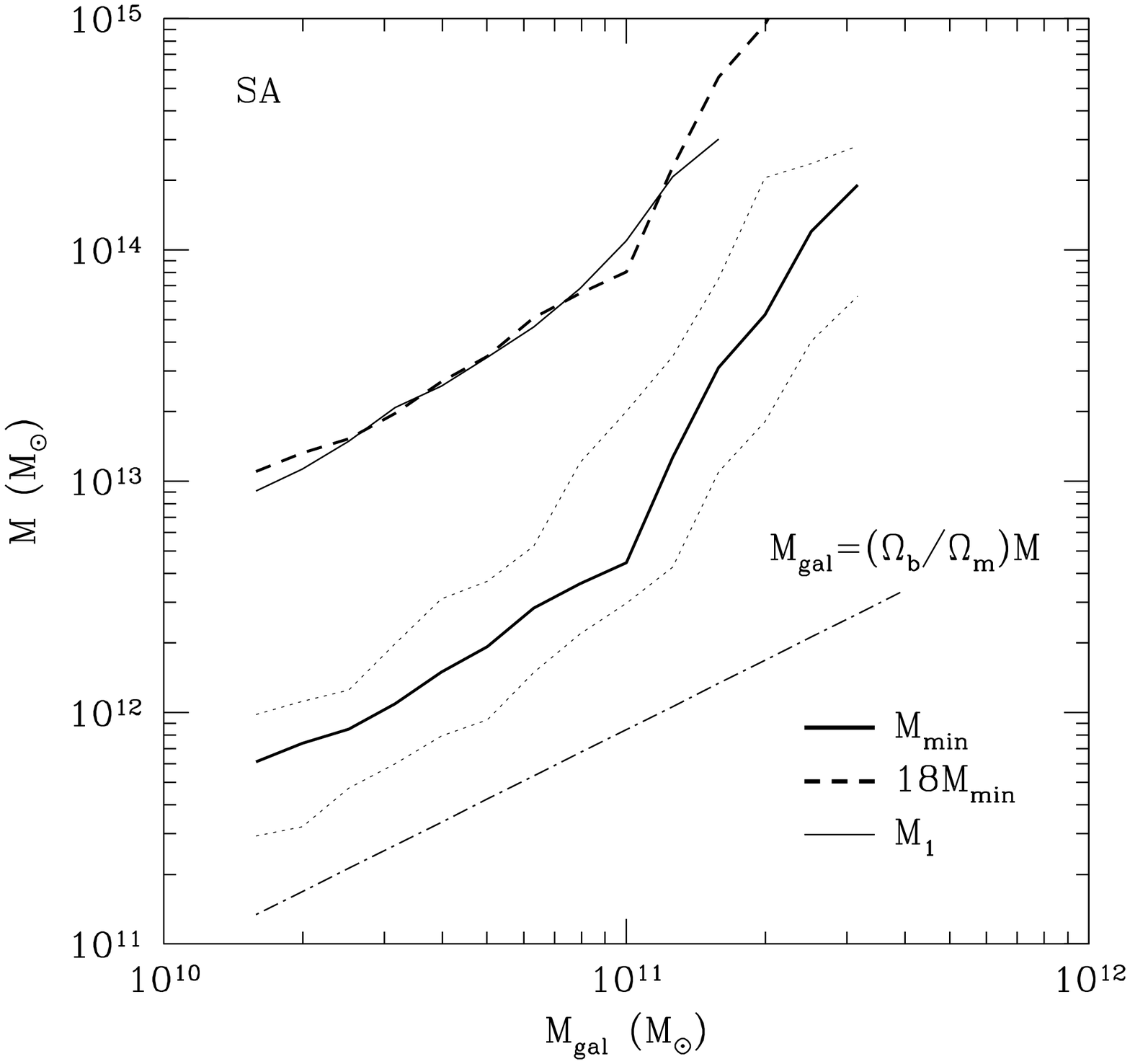} 
\caption[]{\label{fig:MminM1}
HOD parameters as a function of galaxy baryonic mass threshold (left panel
for the SPH simulation and right panel for the SA model). In each panel, 
the thick solid curve is $\Mmin$, the characteristic minimum mass of halos 
that can host galaxies above a given threshold, which is defined here
as the mass of halos that have $\Ncenavg=0.5$. The two dotted curves denotes
masses of halos that have $\Ncenavg=0.1$ and $\Ncenavg=0.9$, respectively.
The thin solid curve is $M_1$, which is the mass of halos that can on 
average host one satellite galaxy above the given threshold. The dashed 
curve indicates 14$\Mmin$ (SPH) or 18$\Mmin$ (SA). For comparison, the 
dot-dashed line shows the baryonic mass corresponding to the universal 
baryon fraction.
}
\end{figure}

\begin{figure}[h]
\epsscale{0.85}
\plotone{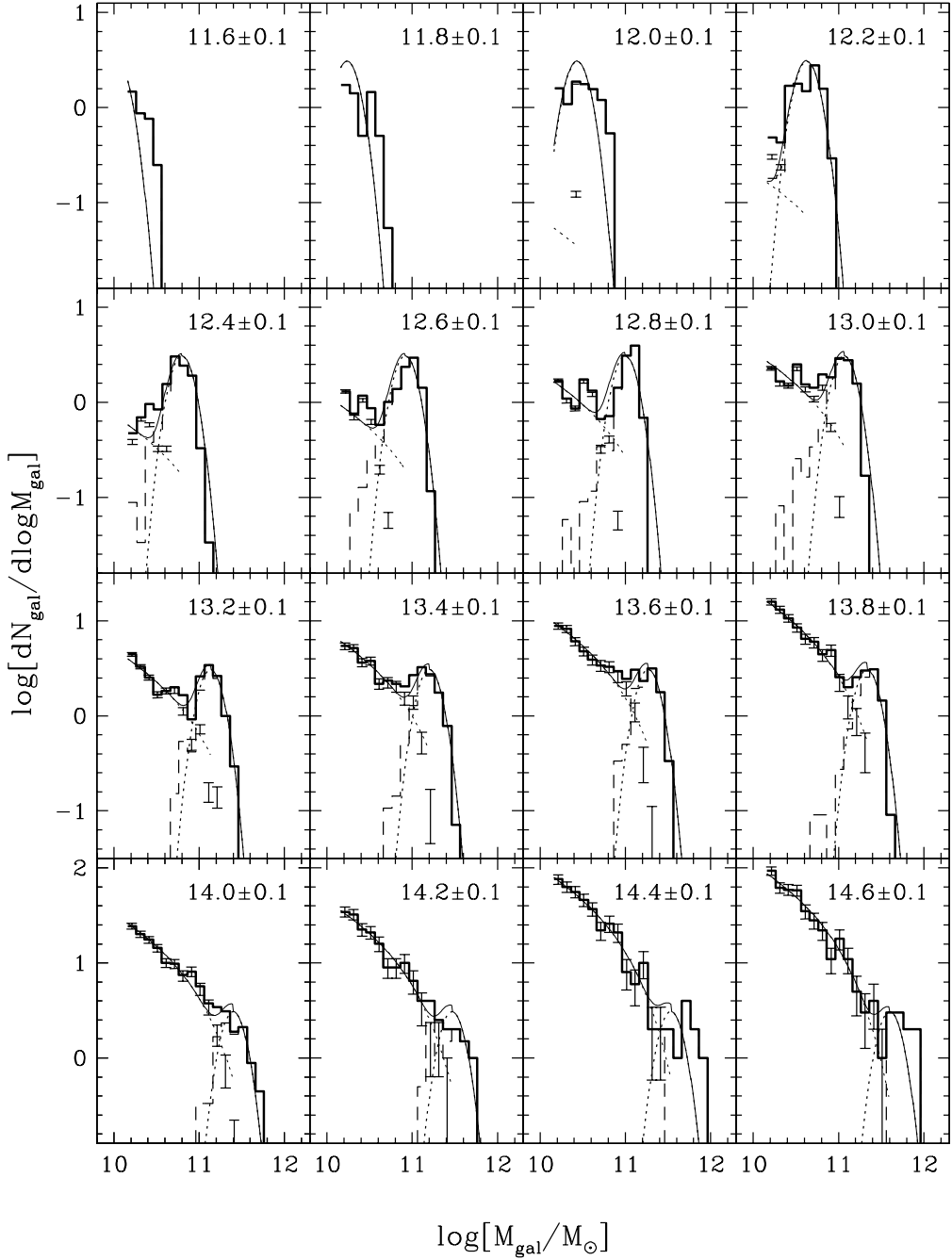} 
\epsscale{1.0}
\caption[]{\label{fig:clf_fit}
Conditional galaxy baryonic mass functions predicted by the SA model as a 
function of 
halo mass (the label in each panel is the range of $\log [M/\Msun]$). In 
each panel,
the CMF is normalized in such a way that the number of central galaxies 
in a halo is unity. The total CMF (thick solid histogram) is decomposed into 
contributions from central galaxies (dashed histogram) and satellites (dots 
with Poisson error bars). The thin solid curve is a sum of a truncated 
Schechter function (dotted curve for satellite galaxies) and a 
Gaussian function (dotted curve for central galaxies), using the parameters 
shown by the dashed lines in the middle and bottom panels of 
Fig.~\ref{fig:clf_fitpar} and further described in the text.
}
\end{figure}

\begin{figure}[h]
\plotone{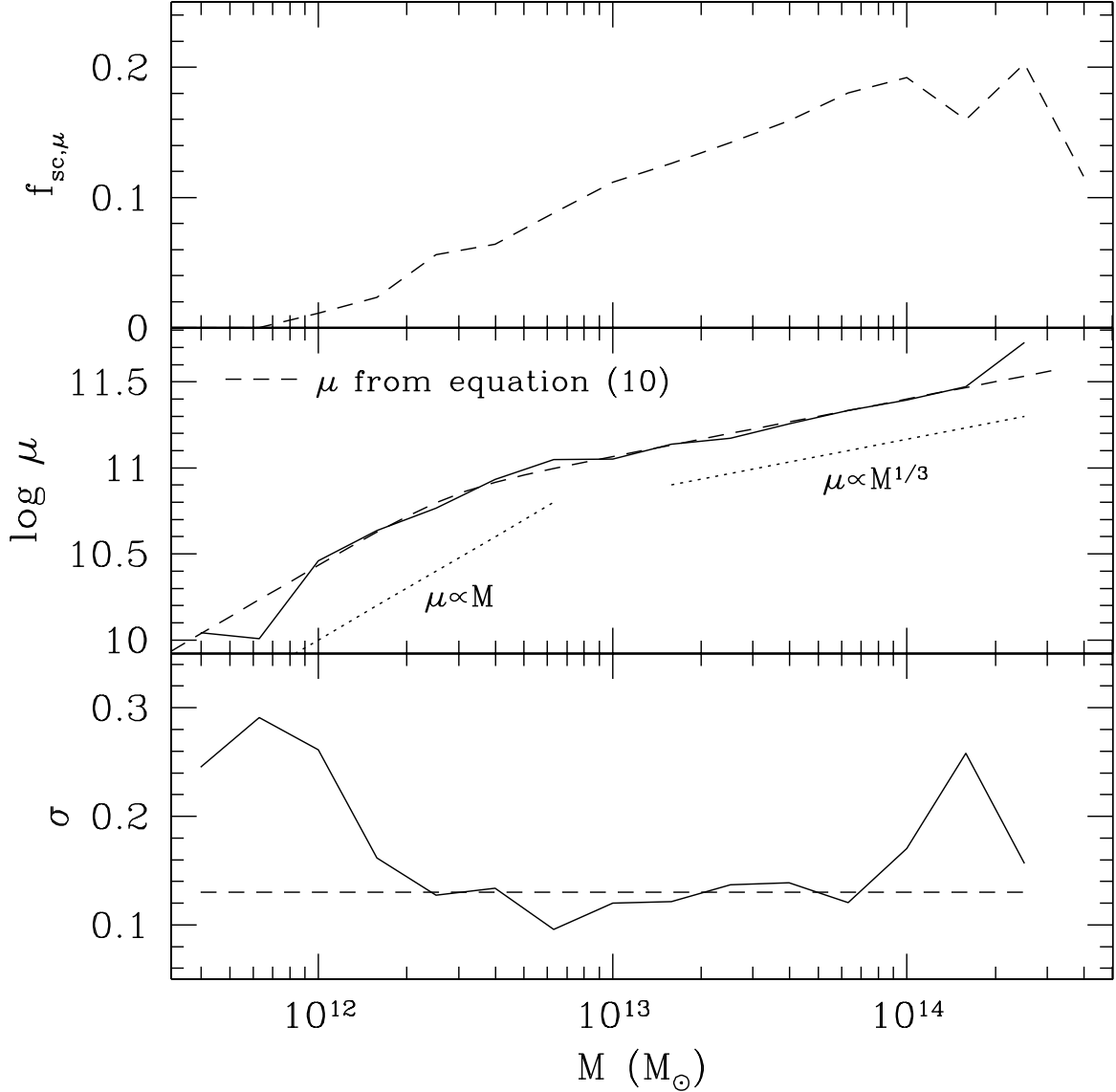}
\caption[]{\label{fig:clf_fitpar}
Parameters related to the CMF in Fig.~\ref{fig:clf_fit}.
The middle and bottom panel show the best-fit values of the mean $\mu$
and the width $\sigma$ of the Gaussian function representing the central 
galaxy CMF. For Gaussian fits shown in Fig.~\ref{fig:clf_fit}, an analytic 
function (dashed curve in the middle panel) is used to represent the mean 
$\mu$, a constant $\sigma$ is adopted (dashed line in the bottom panel),
and the normalization is set to have one central galaxy in a halo.
The top panel shows the ratio of the amplitudes of the truncated 
Schechter function in Fig.~\ref{fig:clf_fit} (CMF for satellite galaxies) 
and the Gaussian function (CMF for central galaxies) evaluated at the mean 
$\mu$ of the Gaussian function.
}
\end{figure}

\begin{figure}[h]
\plotone{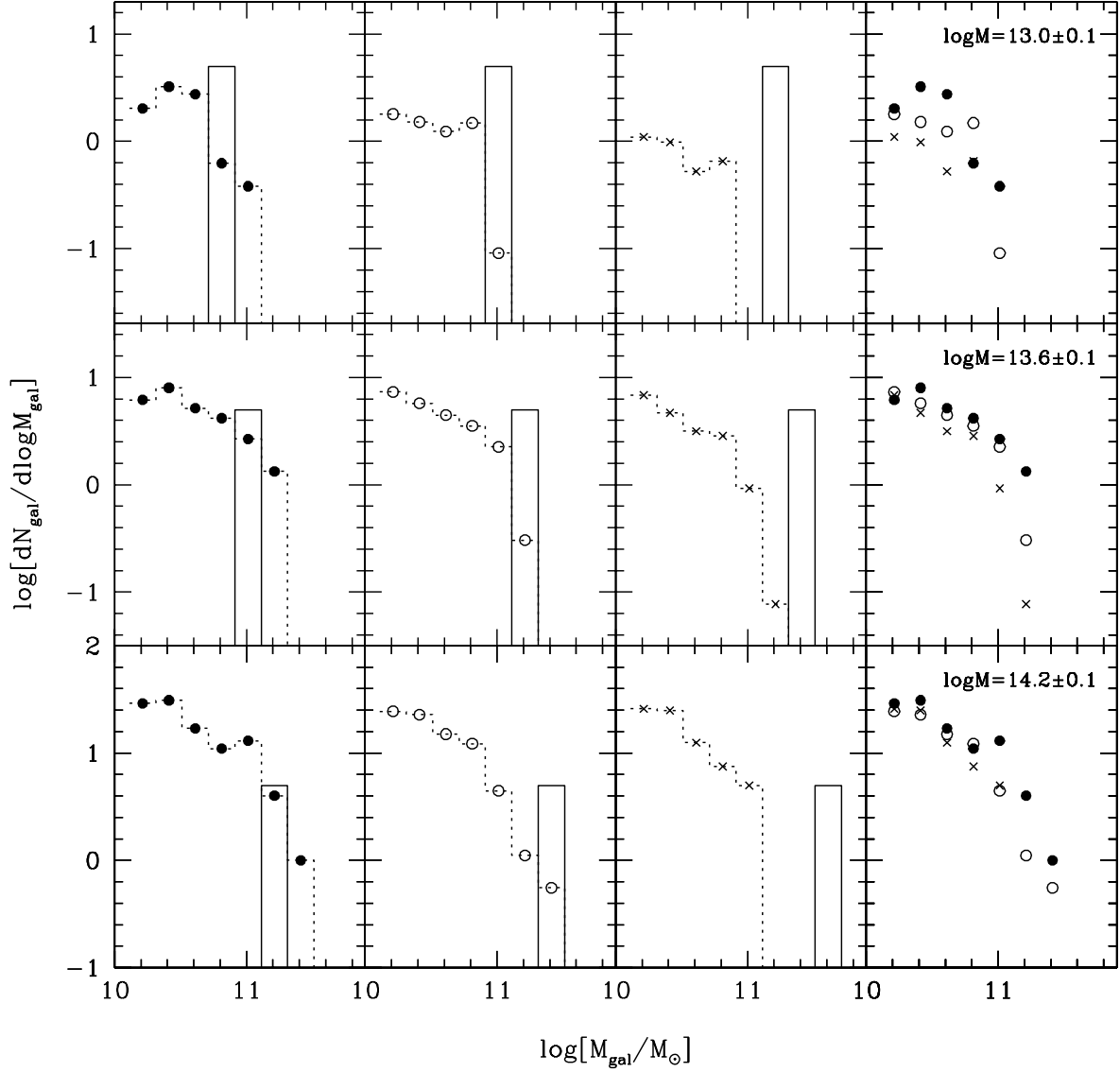}  
\caption[]{\label{fig:ccmf}
The satellite CMF as a function of central galaxy mass in three halo mass bins,
$\log [M/M_\odot]=$13.0 (top), 13.6 (middle), and 14.2 (bottom). For each 
halo mass, the first three panels show satellite CMFs (dotted histograms 
with points) at different central galaxy masses (solid histograms are central 
galaxy CMFs); satellite CMFs in these three panels are plotted together 
in the right panel for direct comparison. CMFs are normalized such that 
there is one central galaxy in a halo.
}
\end{figure}

\begin{figure}[h]
\plotone{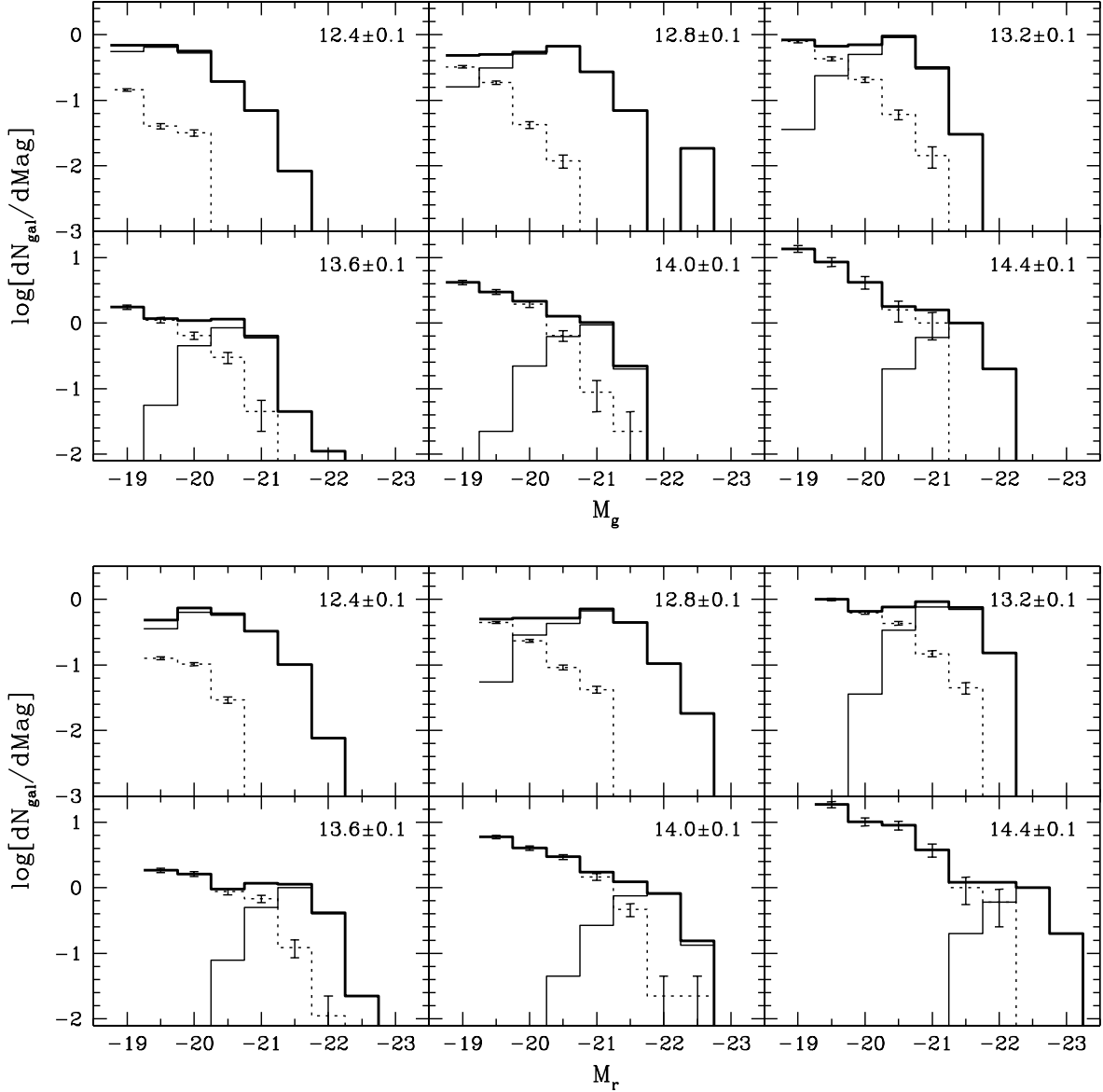}                 
\caption[]{\label{fig:clf_gr}
Conditional luminosity functions in $g$-band (top) and $r$-band (bottom) 
predicted by the SA model as a function of halo mass. Similar to 
Figure~\ref{fig:clf_fit}, the label in each panel marks the range of
$\log [M/M_\odot]$. The dotted, thin solid, and thick solid histograms 
are CLFs for satellite, central, and all galaxies, respectively. The 
faintest bin of the CLF is set by the luminosity above which the completeness 
fraction is unity (see the text).
}
\end{figure}

\begin{figure}[h]
\plottwo{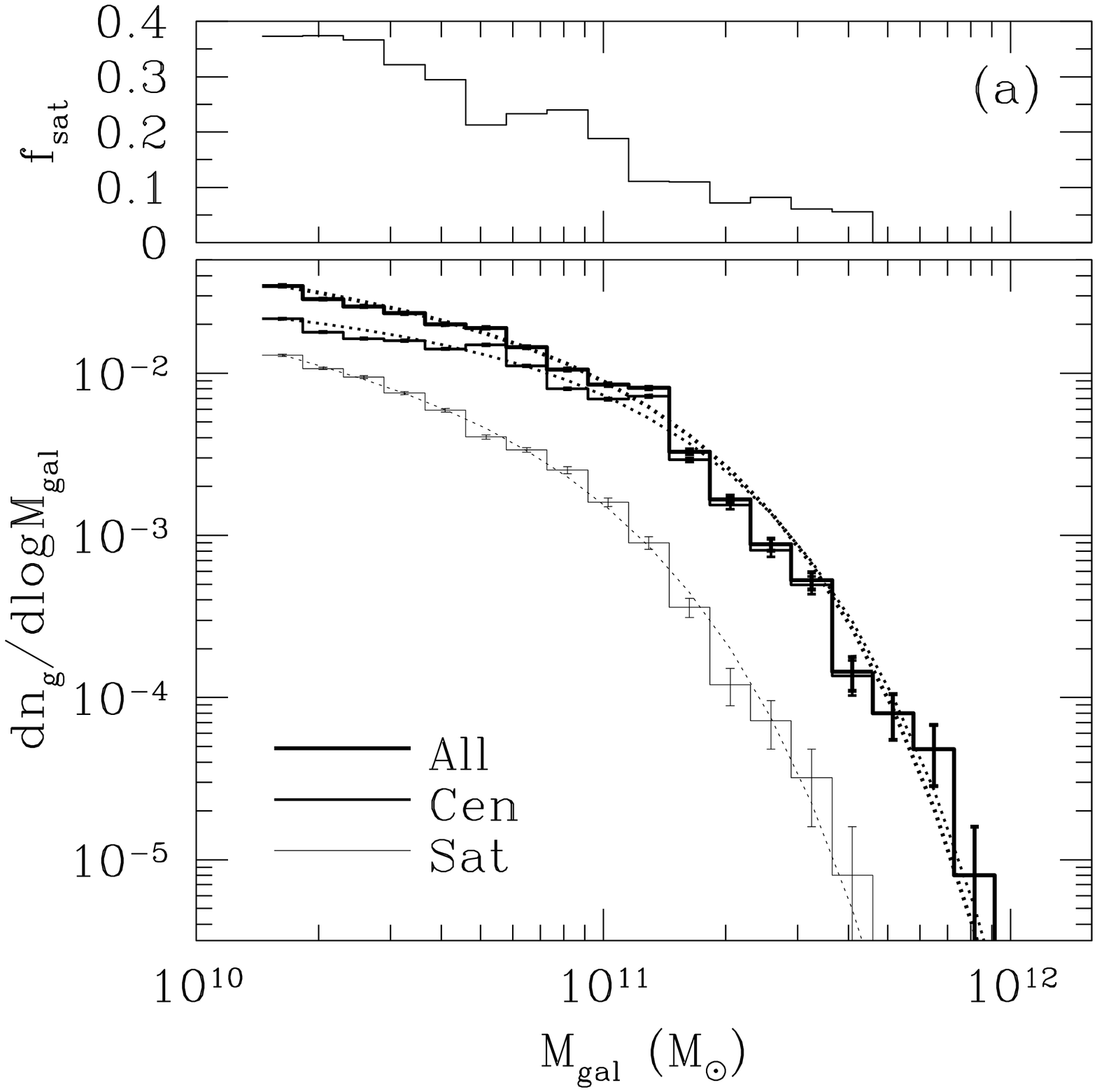}{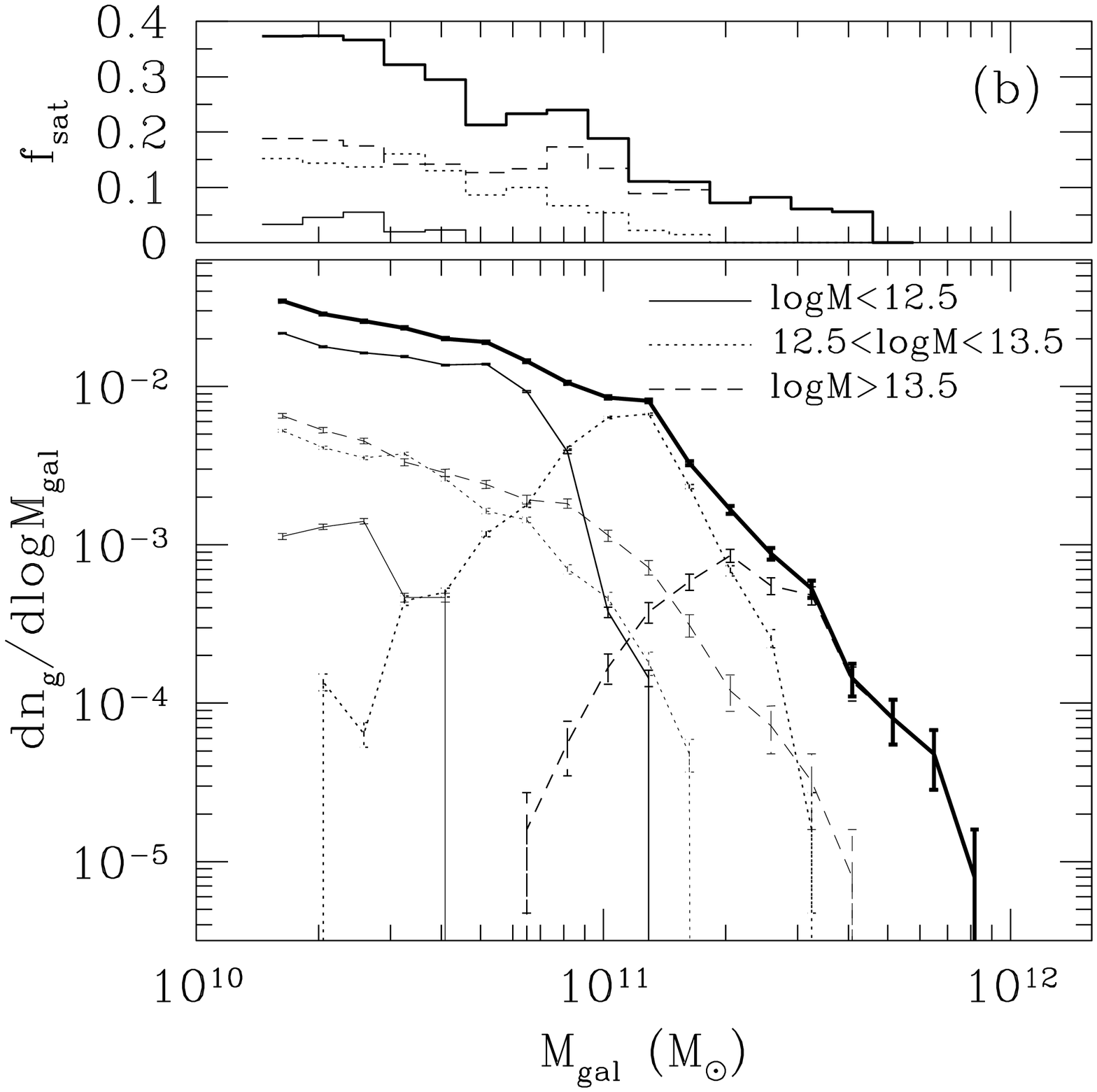}  
\caption[]{\label{fig:lf_cen_sat}
Contributions of central and satellite galaxies to the overall galaxy mass 
function of the SA model. In the lower left panel, the heavy histogram
shows the total mass function (in unit of $h^3{\rm Mpc}^{-3} {\rm dex}^{-1}$),
while lighter histograms show the central and satellite contributions. 
Dotted curves show Schechter function fits. The upper left panel shows 
the fraction of satellite galaxies in each galaxy mass bin. The right panels 
show the separate contributions from different mass ranges, with curves to 
the right showing central galaxy contributions and curves to the left showing 
satellite contributions.
}
\end{figure}

\begin{figure}[h]
\plotone{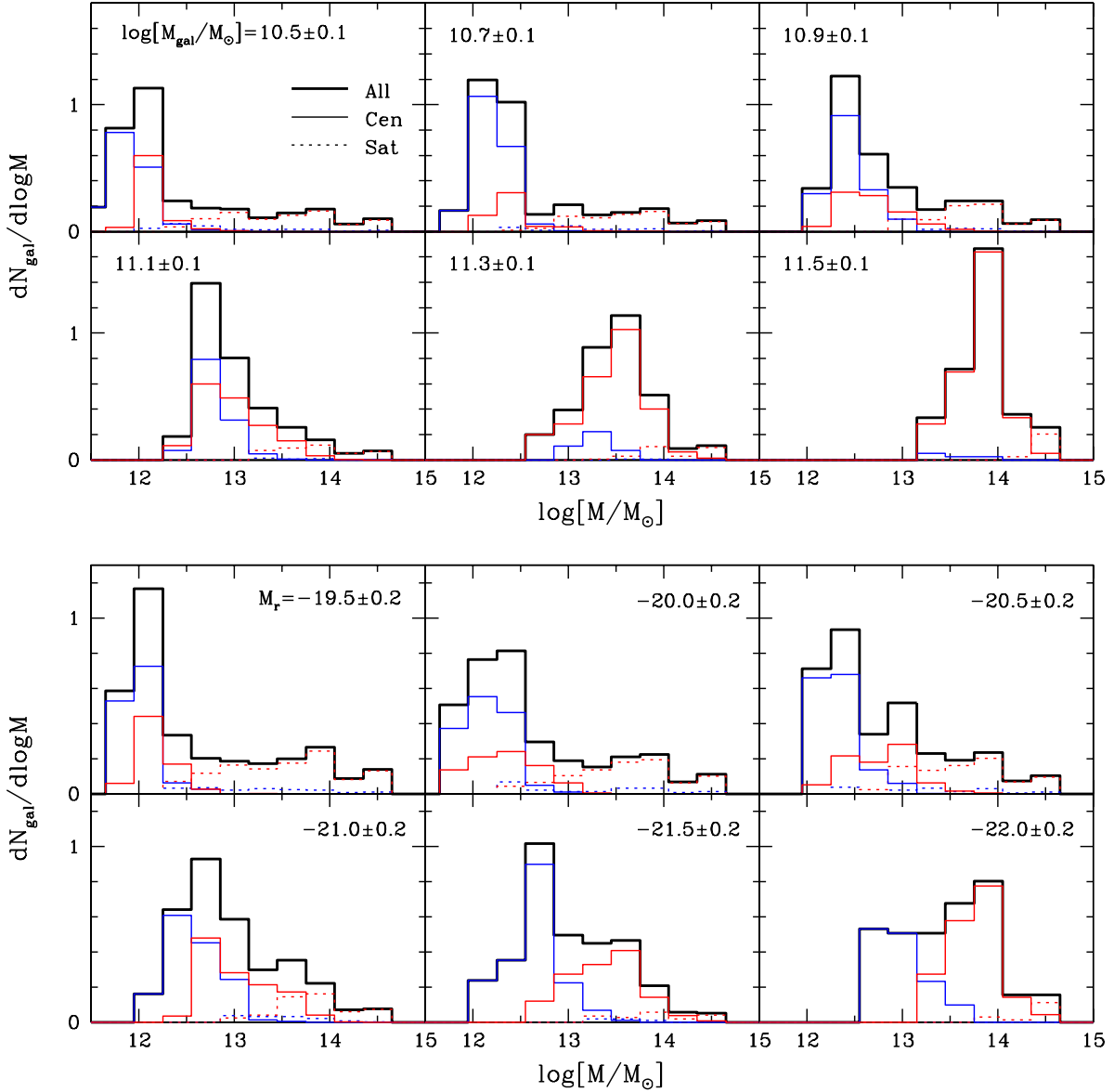}
\caption[]{\label{fig:PML}
Mass distribution of host halos for galaxies at fixed baryonic mass
(top) and at fixed $r$-band luminosity (bottom) predicted by the SA model.
The label in each panel marks the range of $\log [M_{\rm gal}/M_\odot]$
or $M_r$. Contributions from central and satellite galaxies are represented
by thin solid and dotted histograms, respectively. They are further divided
into red and blue histograms for red and blue galaxies with a color cut
at $g-r=0.734$. Distributions are normalized such that the total area under
the thick solid histogram (for all galaxies) is unity.
}
\end{figure}

\end{document}

%% file: table.tex
\begin{deluxetable}{lccccccccc}
\tablewidth{0pt} 
\tablecolumns{10}
\tablecaption{\label{tbl:params} HOD Parameters for Galaxy Samples with
Different Thresholds of Baryonic Mass}
\tablehead{ 
$\ngavg$ & Model & $\log \Mmin$ & $\log M_1$ & $\alpha$ & $\log \Mmin$ & $\sigM$ & $M_0$ & $\log M_1^\prime$ & $\alpha$ }
%\tablecomments{Number density and mass are in units of $h^3{\rm Mpc}^{-3}$ 
%and $\hMsun$, respectively. Columns 2--4 are for the 3-parameter model and 
%Columns 5--9 are for the 5-parameter model (see the text).}
\startdata
0.02   & SPH & 11.67  & 12.96  & 0.97  & 11.68 & 0.15 & 11.86 & 13.00 & 1.02\cr
       & SA  & 11.77  & 12.96  & 1.04  & 11.73 & 0.32 & 12.09 & 12.87 & 0.96\cr
0.01   & SPH & 12.04  & 13.26  & 1.03  & 12.07 & 0.18 & 12.28 & 13.19 & 0.94\cr
       & SA  & 12.03  & 13.30  & 1.02  & 12.02 & 0.26 & 12.28 & 13.32 & 1.07\cr
0.005  & SPH & 12.38  & 13.55  & 1.18  & 12.36 & 0.15 & 12.63 & 13.45 & 1.00\cr
       & SA  & 12.34  & 13.64  & 1.09  & 12.36 & 0.42 & 12.28 & 13.62 & 1.04\cr
0.0025 & SPH & 12.68  & 13.85  & 1.24  & 12.69 & 0.15 & 12.94 & 13.82 & 1.08\cr
       & SA  & 12.58  & 13.91  & 1.16  & 12.60 & 0.28 & 12.77 & 13.86 & 1.03\cr
\enddata
\tablecomments{Number density and mass are in units of $h^3{\rm Mpc}^{-3}$ 
and $\hMsun$, respectively. Columns 3--5 are for the 3-parameter model and 
Columns 6--10 are for the 5-parameter model (see the text). For the 3-parameter 
model, $\Mmin$ is simply set to be the halo mass at which $\Navg=0.5$, and 
$M_1$ and $\alpha$ are obtained through a power-law fit to data points with 
$\Nsatavg>0.1$.}
%\tablenotetext{a}{Number density and mass are in units of $h^3{\rm Mpc}^{-3}$ 
%and $\hMsun$, respectively. Columns 2--4 are for the 3-parameter model and 
%Columns 5--9 are for the 5-parameter model (see the text).}
\end{deluxetable}